\documentclass[letterpaper]{article}
\usepackage[table]{xcolor}
\usepackage[utf8]{inputenc}
\usepackage{authblk}
\usepackage{amsmath}
\usepackage{pgfplotstable}
\usepackage{booktabs}
\usepackage{amsthm}
\usepackage{algorithm}
\usepackage[noend]{algpseudocode}
\newcommand{\vars}{\texttt}

\usepackage[thicklines]{cancel}
\usepackage{amsfonts}
\usepackage{amssymb}
\usepackage{graphicx}
\usepackage{epstopdf}
\usepackage[lmargin=0.8in, rmargin=0.8in, tmargin=1.5in, bmargin=1.2in, headsep=1in]{geometry} 
\usepackage{caption}
\usepackage{multirow}
\usepackage{fancyhdr}
\usepackage[driverfallback=dvipdfm]{hyperref}
\usepackage{subfigure}
\usepackage{pdfpages}
\usepackage[nottoc,notlot,notlof]{tocbibind}
\usepackage{pdflscape}
\usepackage{amsmath, xparse}
\usepackage{listings}
\usepackage{enumerate}
\usepackage{multirow}
\usepackage{lscape}
\usepackage{rotate}
\usepackage{longtable}
\usepackage{rotating}
\usepackage{amsbsy}
\usepackage{footnote}
\usepackage{enumitem}
\usepackage{setspace}
\usepackage{array}
\usepackage{color}
\usepackage{colortbl}
\usepackage{trimclip}
\usepackage{tcolorbox}
\usepackage{blindtext}
\theoremstyle{definition}

\usepackage{gensymb}
\newcommand{\vecl}[1]{\mathbf{#1}} 

\date{\today}

\definecolor{flightmechanics}{rgb}{1,1,0.7}
\definecolor{flightmechanics1}{rgb}{1,1,0.85}
\definecolor{biomechanics}{rgb}{0.65,0.85,0.9}
\definecolor{biomechanics1}{rgb}{0.65,0.83,0.8}
\definecolor{aerodynamics}{rgb}{1,0.6,0.5}
\definecolor{aerodynamics1}{rgb}{1,0.55,0.45}
\definecolor{control}{rgb}{0.31,0.75,0.5}
\definecolor{control1}{rgb}{0.31,0.7,0.47}
\definecolor{header}{rgb}{1,0.85,0}
\definecolor{raw}{rgb}{1,1,0.7}
\definecolor{raw1}{rgb}{1,1,0.85}
\makeatletter
\renewcommand*\env@matrix[1][\arraystretch]{%
  \edef\arraystretch{#1}%
  \hskip -\arraycolsep
\makeatother
  \let\@ifnextchar\new@ifnextchar
  \array{*\c@MaxMatrixCols c}}

\title{Stability and sensitivity analysis of bird flapping flight}
\author[1]{Gianmarco Ducci}
\author[1]{Victor Colognesi}
\author[1]{Gennaro Vitucci}
\author[1]{Philippe Chatelain}
\author[1]{Renaud Ronsse}
\affil[1]{Institute of Mechanics, Materials, and Civil Engineering; UCLouvain; Louvain-la-Neuve, Belgium}
\date{}
\begin{document}
\maketitle
\begin{abstract}
This paper investigates stability analysis of flapping flight. Due to time-varying aerodynamic forces, such systems do not display fixed points of equilibrium. The problem is therefore approached via a limit cycle analysis based on Floquet theory. Stability is assessed from the eigenvalues of the Jacobian matrix associated to the limit cycle, also known as the Floquet multipliers.\\ 
We developed this framework to analyze the flapping flight equations of motion of a bird in the longitudinal plane. Such a system is known to be not only non-linear and time-dependent, but also driven by state-dependent forcing aerodynamic forces. A model accounting for wing morphing under prescribed kinematics is developed for generating realistic state-dependent aerodynamic forces. The morphing wing geometry results from the envelope of continuously articulated rigid bodies, modeling bones and feather rachises, and capturing biologically relevant degrees of freedom. A sensitivity analysis is carried out which allows studying several flight configurations in trimmed state. \\
Our numerical results show that in such a system one instability mode is ubiquitous, thus suggesting the importance of sensory feedback to achieve steady-state flapping flight in birds. The effect of wingbeat amplitude, governed by the shoulder joint, is found to be crucial in tuning the gait towards level flight, but marginally affects stability. In contrast, the relative position between the wing and the center of mass is found to significantly affect the values of Floquet multipliers, suggesting that the distribution of pitching moment plays a very important role in flapping flight stability. \\

\end{abstract}
\newpage
\section{Introduction}
Biological fliers have been a source of scientific inspiration for decades, thanks to their impressive performance. Due to a recent interest in flapping vehicles, there is a strong effort from the scientific community to unveil the bio-mechanics of animal flapping flight.\\
Control capabilities, in particular, are outstanding as demonstrated by accelerations of up to 14G, roll rates up to $5000\frac{deg}{s}$ achieved by barn swallows (\textit{Hirundo rustica)}~\cite{shyy1999}, and the ability of mitigating environmental perturbations such as wind and gust. It is therefore critical to establish flight dynamic stability in order to investigate the mechanisms governing such behaviors. Over the years, a lot of work has been accomplished to assess flight dynamic stability at different scales.\\
The first attempt to assess the longitudinal stability in flapping flight was carried out by Taylor and Thomas~\cite{taylor2002}. They addressed the problem using a quasi-static and blade element approach in order to estimate aerodynamic loads. Subsequently they analytically evaluated the static stability, by considering the variation of the pitching moment with respect to the angle of attack. They concluded that, in flapping flight, longitudinal stability drastically depends on where the quasi-static flight force acts with respect to the body center of mass. They suggested that particular wing motions, such as sweeping, has significant impact on the overall system stability.\\
However, static stability is only a necessary condition for dynamic stability. Early studies investigating dynamic stability leveraged on averaging the system dynamics over the flapping period. In particular,~\cite{taylor2003} and~\cite{xiong2008}, linearized the equations of motion from experimental measurements of the aerodynamic derivatives.
The averaging approach however fails if the wingbeat frequency is close to the natural frequency of the body motion, such as for large birds in slow forward flight, or in transitions between two different flight regimes, such as from fast forward flight to hovering~\cite{taylor2002, iosilevskii2014b}.\\
Taylor and \.{Z}bikowski~\cite{taylor2005} conducted the first study of stability in terms of periodic orbits, re-defining the stability of flapping flight as the asymptotic orbital stability in the phase space. For the first time, they introduced a limit cycle approach to study flapping flight dynamics.\\
In~\cite{dietl2008}, Dietl and Garcia followed this approach by introducing forcing aerodynamic terms. They defined the trim condition as the limit cycle described by the state-space variables of the equations of motion with the same period as the flapping wingbeat, and used Floquet theory to determine its stability. They studied the longitudinal flight dynamics of an ornithopter treated as a rigid body, with imposed joint kinematic trajectories, and developed a limit-cycle detection method based on a multiple-shooting algorithm to concomitantly identify the limit-cycle, and assess its stability. Importantly, they restricted the kinematic analysis to two degrees of freedom only, namely the plunging angle, and the wing twist, defining the wingbeat amplitude and the angle of attack respectively. This model did not capture important features of the wing, such as morphing and sweep angle.\\
From a bird perspective, the wing kinematics also defines the flight regime. Indeed, birds can modulate different degrees of freedom in order to adapt their forward speed, gain or lose altitude, and perform maneuvers. To the best of our knowledge, an extensive analysis of the influence of such parameters on the stability properties of the flier, and on its performance, has not been conducted to date.\\
Based on the aforementioned definition of trimmed flight, this paper reports a method that relies on a multiple-shooting algorithm to identify limit cycles for large birds dynamics and evaluate their stability via Floquet theory. In particular, we couple an aerodynamics model capable of dealing with poly-articulated morphing wings to the bird body moving in space. Finally, we identify a family of limit cycles corresponding to different flight regimes (climbing, descending and level flight).\\
Furthermore, this analysis is successfully employed to investigate the sensitivity of flight stability to the reciprocal position between the wing root and the bird center of mass.\\
Two impacting results have been found. First, the wingbeat amplitude defines the flight regime (climbing, descending or level) without significantly affecting its stability. Second, the wing insert position drastically impacts the system stability, suggesting that the attitude of the wing for developing nose-down moment is beneficial in terms of longitudinal stability.\\
 The rest of the paper is structured as follows. Section~\ref{sec:dynamics_aerodynamics} reports the dynamic model of the bird, and more particularly the coupling between the dynamical model of the animal and the aerodynamic loading. Section~\ref{sec:methods} presents the multiple-shooting algorithm and the numerical parameters employed in the simulations. Section~\ref{results} reports the results obtained in the numerical investigations. Finally, in Section~\ref{sec:conclusions}, we discuss the influence the wingbeat amplitude and wing insert position have on the dynamic stability of the flier. The paper is finally concluded and some perspectives for controlling and achieving stable flight inspired by our model, are outlined.

\section{Dynamical model of a flying bird}\label{sec:dynamics_aerodynamics}
In this section, the equations of motion of a flying bird are developed. We build this model upon three main assumptions:
\begin{enumerate}
\item The flight is restricted to the longitudinal plane, so that the bird main body has only 3 degrees of freedom: 2 in translation and 1 in rotation. The system is symmetric with respect to this plane. As a consequence, lateral forces, rolling moments and yawing moments are identically equal to zero at every time and therefore do not have to be considered in the equations of motion.  Stability of these degrees of freedom is thus not discussed in the present paper.
\item The inertial effect of the wings on the main body can be neglected. The model therefore does not account of the effect of flapping on the motion of the center of gravity. This is guided by the fact that for large and fast migratory birds, the wing mass is much lower than the body mass, about 5\% according to~\cite{vanderberg1995}. This assumption has been extensively used for both ornithopter and insect scale models~\cite{taylor2005, dietl2008, dietl2011, taha2014}. 
\item The aerodynamic effects due to the tail are not explicitly modeled. The proposed model compensates for the missing tail by sweeping of the wing around the body center of mass, thus providing the possibility to generate both nose-up and nose-down pitching moments.
\end{enumerate} The main morphological parameters used to model the main body and the wing kinematics are introduced in the following sections.

\subsection{Equations of longitudinal motion}
\begin{figure}[htbp]
\centering
\includegraphics[width=.50\textwidth]{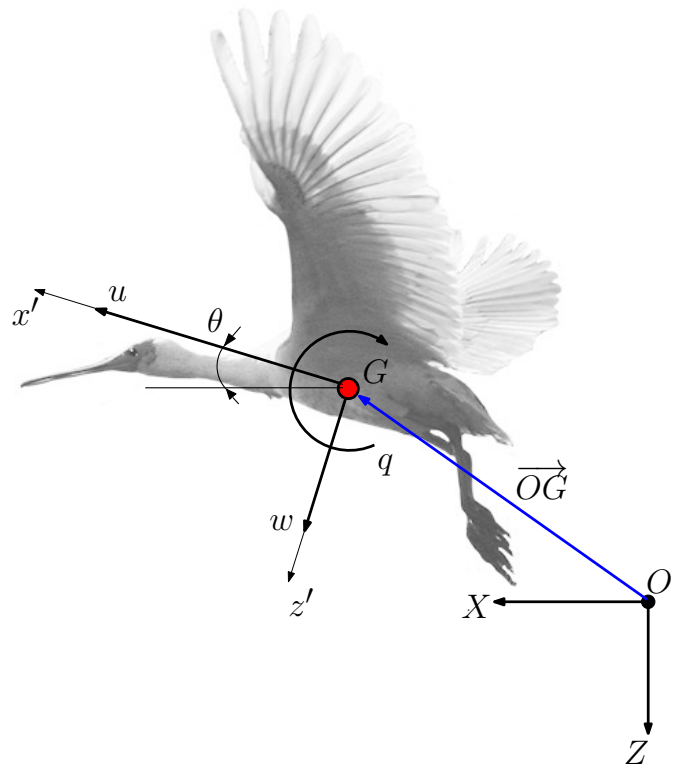}
	\caption{Reference frames describing flight dynamics in the longitudinal plane. The origin of the moving body-frame is taken at the bird's center of mass $G$. Image re-adapted from \url{www.fotki.com}.}
\label{fig:birdframe}
\end{figure}

The considered bird is modeled as a rigid body of mass $m_{b}$ located at the center of mass $G$, and moving in the longitudinal plane. Two reference frames are required in order to study the longitudinal flight dynamics, as shown in Figure \ref{fig:birdframe}: a fixed inertial frame $O(X, Y, Z)$, and a moving body frame $G(x', y', z')$ with unit vectors $(\hat{\vecl{e}}_{x'}, \hat{\vecl{e}}_{y'}, \hat{\vecl{e}}_{z'})$. Since only longitudinal motions are considered, the body state is captured by three degrees of freedom: the translations along the $X$ and $Z$ axes, and the relative angle between both frames, i.e. the so-called pitch angle $\theta$~\cite{etkin1959}. The body frame is assumed to be centered in the center of mass of the bird, and oriented by taking the $x'$-axis pointing forwards, the $z'$-axis pointing downwards, and $y'$-axis to define a right-handed frame (Figure~\ref{fig:birdframe}). \\ 

Under the aforementioned hypotheses, the set of Newton-Euler equations governing the longitudinal dynamics is obtained following the conventional equations for fixed-wing aircraft, i.e. in the form of~\cite{etkin1959, casarosa2013}
\begin{equation}
\begin{aligned}
	\dot{u} &= -qw - g\sin \theta + \frac{1}{m_b}{F_{x'}(\mathbf{x}(t),  t)}\\ 
	\dot{w} &= qu + g\cos \theta +\frac{1}{m_b} F_{z'}(\mathbf{x}(t),  t)\\
	\dot{q} &=\frac{1}{I_{yy}}M_{y'}(\mathbf{x}(t),  t)\\
	\dot{\theta} &= q
\label{eqn:eom2}
\end{aligned}
\end{equation}
where $u$ is the velocity component along local axis $x'$, $w$ is the velocity component along local axis $z'$, $\theta$ is the pitch angle describing the body inertial orientation, $q$ is its derivative, i.e. the body angular velocity, positive according to the orientation given in Figure~\ref{fig:birdframe} and $t$ is the time. The parameter $I_{yy}$ is the moment of inertia about the $y'$-axis.\\Accordingly, the state vector describing the longitudinal motion is
\begin{equation*}
\mathbf{x} = \left \{ u, w, q, \theta \right \}
\label{eqn:statevariables}
\end{equation*}

The forcing terms in Equations~(\ref{eqn:eom2}), namely $F_{x'}$, $F_{z'}$, and $M_{y'}$, are the aerodynamic loads acting on the bird. Consequently, the bird model is actually a 4-states \textit{non-autonomous} system, where the aerodynamic terms at the generic time $t>0$ depends not only on the instantaneous state vector, but also on the instantaneous configuration of the wing $\varphi(t)$ in the flapping cycle. This is obviously the main difference with respect to an equivalent fixed-wing aircraft model.\\
Formally, the forcing terms depend on the whole past state history. Therefore, these forcing aerodynamic terms can be expressed in the form of Equation~(\ref{eqn:eom2}) only if a quasi-steady-state approximation is used, which typically holds for large scale birds~\cite{taha2012}.

\subsection{Wing kinematics}
\begin{figure}[!h]
\begin{center}
\includegraphics[scale=.5]{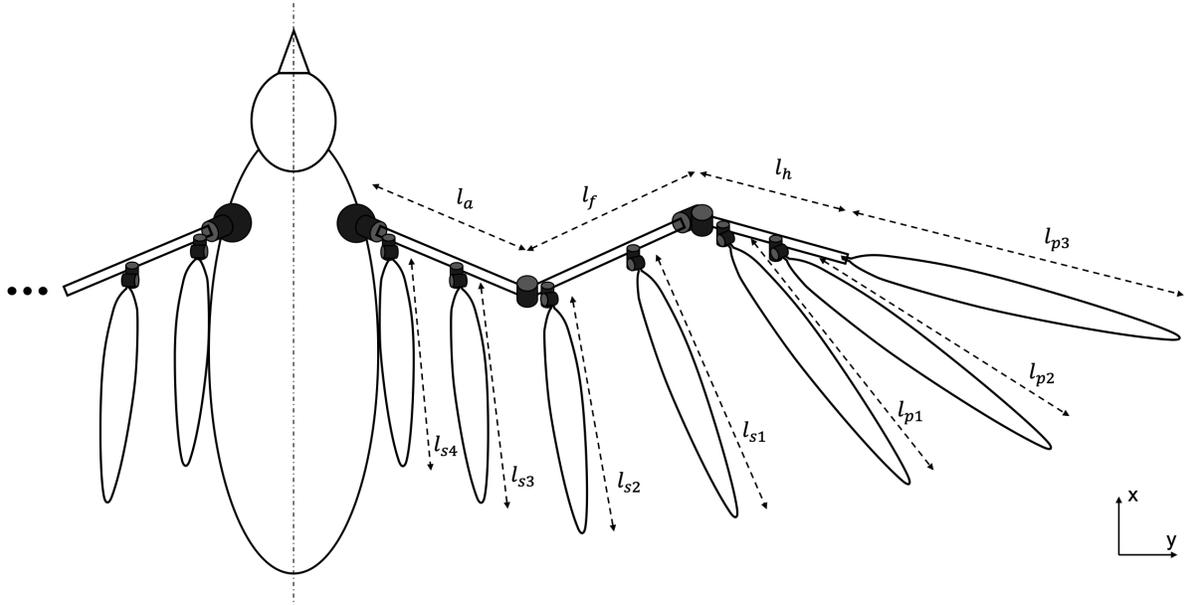}
\caption{Top view of the bird wing kinematic model. Each cylinder captures a rotational degree of freedom, i.e. three at the shoulder (between the body and arm), one at the elbow (between the arm and forearm), and two at the wrist (between the forearm and hand). Moreover, each master feather is attached to one of these bodies via two more degrees of freedom, except the most distal one which aligned with the last bone segment.}
\label{fig:wingbodies}
\end{center}
\end{figure}

The wings are attached to the main body and feature kinematic degrees-of-freedom as pictured in Figure~\ref{fig:wingbodies}. They both consist of three rigid bodies corresponding to the bird arm, forearm and hand. Relative motions between these segments govern wing morphing. To these wing segments are attached "master-feather" bodies representing the plumage and capturing its movements.

We do not solve the wing dynamics in the state space equations of the flier, but we rather assume that their kinematics are imposed. Consequently, the internal torques in the wing joints do not have to be computed for solving the body equations of motion. The description of the right and the left wing kinematics are assumed to be mirror, since movements are symmetric. For the sake of simplicity, each joint angle $i$ is considered to follow a harmonic trajectory $q_i(t)$, with respective amplitude $A_i$, offset $q_{0,i}$ and phase $\phi_{0,i}$:
\begin{equation}
q_i(t) = q_{0,i} + A_i \sin\left(\omega t + \phi_{0,i}\right)
\label{eqn:general_joint}
\end{equation}
with $\omega = 2\pi f$ and $f$ is the flapping frequency, identical for each joint. For the six rotational joints of the model in Figure \ref{fig:wingbodies}, this makes a total of 19 gait parameters (including the wingbeat frequency) prescribing a particular set of wing kinematics. \\
Feather movements are governed by a simplified version of the model developed in~\cite{colognesi2020}, which is itself inspired from~\cite{Wu:2003}. Indeed, feathers are similarly attached to the wing bodies via two rotational degrees of freedom (allowing spreading and pitching  in a bone-relative frame of reference) but the motion of these degrees of freedom follows here predefined trajectories, while they feature some dynamic compliance in~\cite{colognesi2020, Wu:2003}. More precisely, we constrained feather movements via kinematic relationships depending on the angles between the wing segments in order to make them spreading and folding smoothly with the wing. These kinematic relationships are reported in the online version of the code available at \url{https://github.com/vortexlab-uclouvain/multiflap}.
\subsection{Aerodynamic model}
\begin{figure}[!htbp]
\begin{center}
\begin{minipage}[c]{.45\textwidth}
\centering
\includegraphics[width=1.\textwidth]{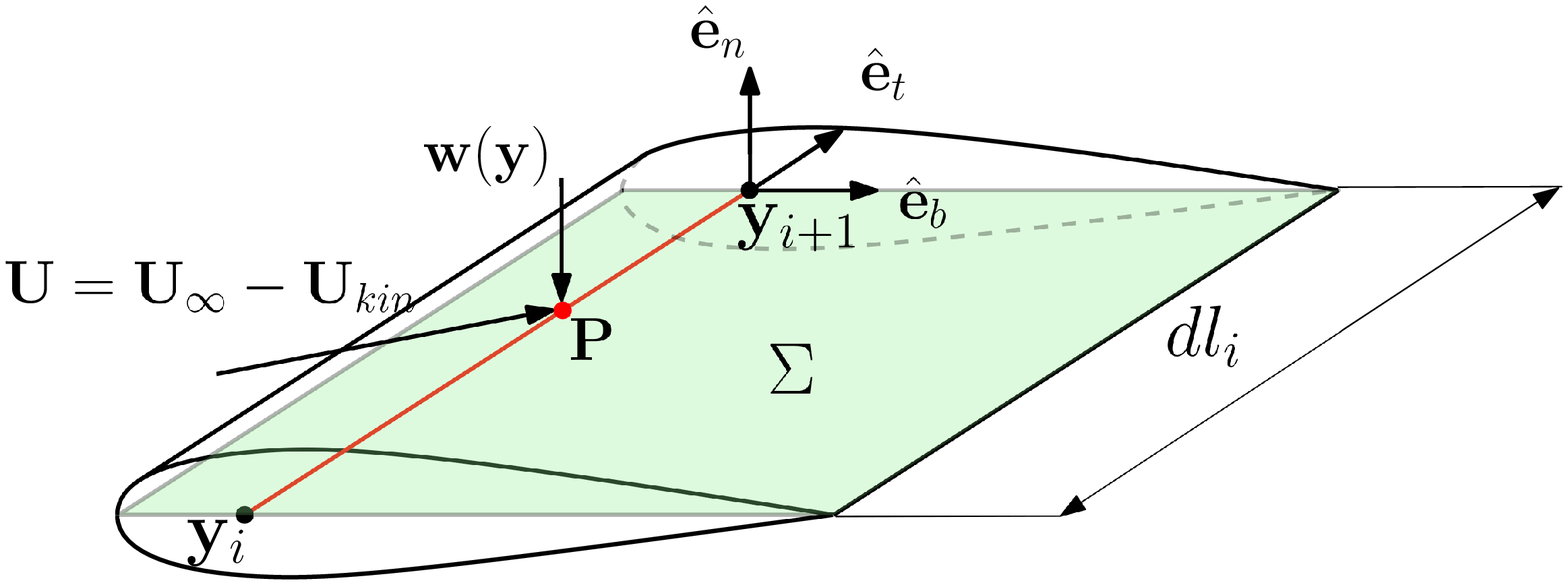}
\end{minipage}%
\hspace{10mm}%
\begin{minipage}[c]{.45\textwidth}
\centering
\includegraphics[width=1.\textwidth]{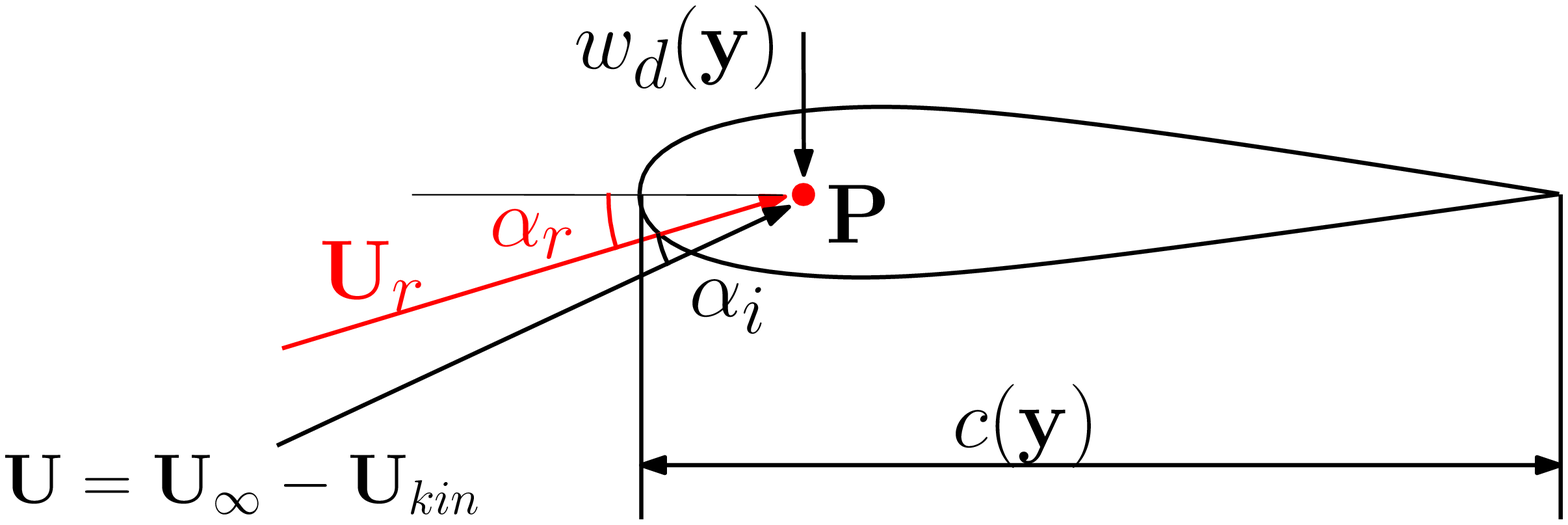}
\end{minipage}
\end{center}
	\caption{Left: Wing element between two wing profiles, and identifying a plane $\Sigma$ containing the lifting line. Right: Cross section containing the chord point $\vecl{P}$ where the velocities are applied.}
\label{fig:profile_velocities}
\end{figure}

In this section, the model used to compute the aerodynamic forces acting on the wing is developed. The model assumes that all aerodynamic forces act on the wings, and none on the main body. We use a quasi-steady lifting line approach, where the wake is shed backward in the form of straight and infinitely long vortex filaments at each time-step of the simulation.  More details about this aerodynamic model are reported in~\cite{colognesi2020}.
\begin{figure}[!htbp]
\centering
\includegraphics[width=.45\textwidth]{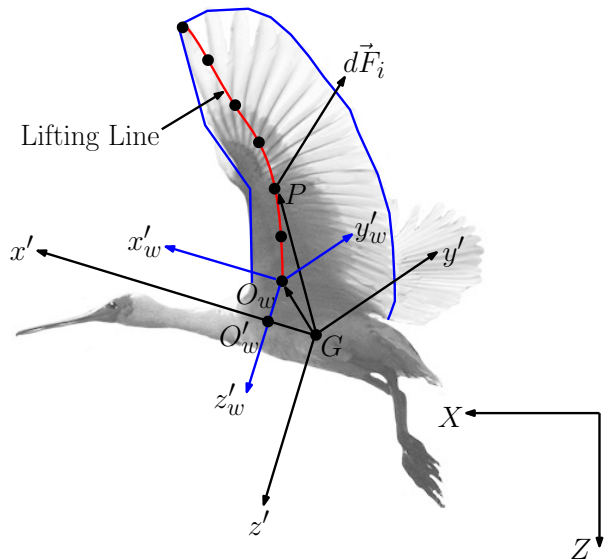}
\caption{ Aerodynamic forces acting on the wing are applied at the discretized points of the lifting line (red), over the wing span. Adapted from \url{www.fotki.com}.}
\label{fig:liftinglinemodel}
\end{figure}
The wing motion and its position are defined in a wing-bone frame $(x_{w}, y_{w}, z_{w})$ shown in Figure~\ref{fig:liftinglinemodel}. The respectives unit vectors along these axes are $(\hat{\vecl{e}}_{x'_{w}}, \hat{\vecl{e}}_{y'_{w}}, \hat{\vecl{e}}_{z'_{w}})$. This frame is taken to follow the orientation of the body frame, while the translation of its origin determines the position of the bird shoulder with respect to its center of mass. The projection of $O_{w}$ on the $x'$ axis, identifies the $O'_{w}$ point.
From this wing position, the lifting line is then consequently extracted. It is defined as the line passing through the quarter of the chord, which is itself defined as the segment orthogonal to the lifting line, going from the leading to the trailing edge of the wing. From a given wing configuration, the leading and trailing edges are defined as follows. The former goes from the shoulder to the wrist in a straight line, then to the tip of the outermost feather. The latter connects the tip of each feather from the innermost to the outermost. The lifting line is then obtained through an iterative process guaranteeing that it is located at the quarter of the chord distance and that it is orthogonal to the chord at each points.\\
In order to compute the aerodynamic forces, it is required to know the wing angle of attack. A generic wing cross section is shown in Figure~\ref{fig:profile_velocities}, where $c(\vecl{y})$ represents the aerodynamic chord lenght. Each wing element is identified by a plane containing the lifting line. The unit vector orthogonal to such a plane is denoted by $\hat{\vecl{e}}_n$, the unit vector tangent to the lifting line $\hat{\vecl{e}}_t$ and the binormal one $\hat{\vecl{e}}_b=\hat{\vecl{e}}_t \times \hat{\vecl{e}}_n$. \\
According to this notation, $-w_{d}\hat{\vecl{e}}_{n}$ is the induced velocity (downwash) and $\vecl{U} = \vecl{U}_{\infty} - \vecl{U}_{kin} $ is the relative velocity seen by a wing profile, which accounts for the flight speed $\vecl{U}_{\infty}$ and the wing motion $\vecl{U}_{kin}$ while its component along $\hat{\vecl{e}}_t$ is previously eliminated.\\
Hence, the effective angle of attack is given by

 \begin{equation}\label{eq:alpha_ind}
\alpha_{r} = \alpha - \alpha_{i}\simeq \alpha -  \frac{w_{d}}{|\vecl{U}|}
\end{equation}

\begin{figure}[!h]
\begin{center}
\includegraphics[scale=.4]{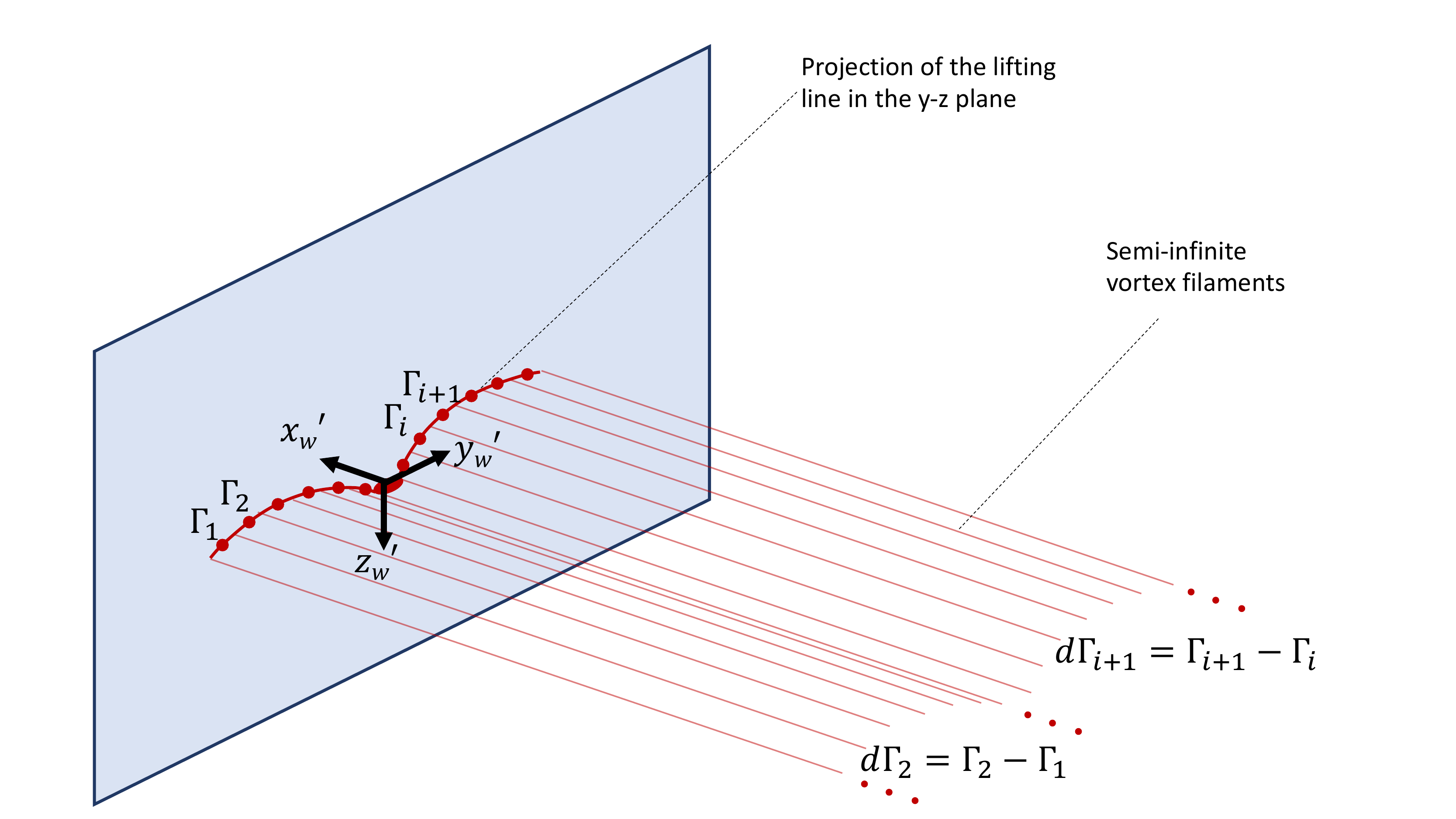}
\caption{The vortex wake of the bird is considered straight and infinite at each time-step of the flapping period. The variations in the line circulation $\Gamma$ induce vortex tubes of circulation $d\Gamma_{i+1} = \Gamma_{i+1}-\Gamma_i$, which in turn induce velocities in the wake.}
\label{fig:lifting_line_wake}
\end{center}
\end{figure}

The wake is considered to be composed of semi-infinite vortex tubes aligned with the x-axis, as shown in Figure~\ref{fig:lifting_line_wake}. In theory, because the wing is not straight, the bound vortex (i.e. the circulation of the lifting line itself) also induces velocities on the line itself. However in the presented model, these induced velocities are neglected since their magnitude is much lower as compared to the flight velocity. Therefore the only induced velocity accounted for is a vertical component due to the wake. This velocity at a point $\vecl{y}\equiv(y'_{w}, z'_{w}) $ induced by a set of semi-infinite vortex tube of circulation $d\Gamma_{i}$ is computed via Biot-Savart law~\cite{buresti2012}, i.e.
\begin{equation}\label{eq:integration_BS}
	w_{d}(\vecl{y}) = -\frac{1}{4\pi}\sum_{i}\Big(-d\Gamma_{i}\dfrac{(\vecl{y}-\vecl{y}_{i})\times\hat{\vecl{e}}_{x'_{w}}}{\left|\vecl{y}-\vecl{y}_{i}\right|^2}\Big) \cdot \hat{\vecl{e}}_{n} 
\end{equation}
where $i$ are the discretized elements of the lifting line.\\
Considering the theorem of Kutta-Joukowski, the local circulation $\Gamma$ is computed as
\begin{equation}\label{eq:circulation_zero}
\Gamma(\vecl{y}) = \frac{1}{2}|\vecl{U_{r}}(\vecl{y})|c(\vecl{y})C_{l\alpha}(\alpha - \alpha_{i})
\end{equation}
where $\vecl{U_{r}}$ is the norm of the local relative velocity vector, $c$ is the local chord, and $C_{l\alpha}$ is equal to $2 \pi$ as a result of thin airfoil theory.
We now assume the downwash velocity to be sufficiently small to approximate $|\vecl{U_{r}}|(y)\simeq |\vecl{U}|$. Considering Equation~(\ref{eq:circulation_zero}) and injecting Equation~(\ref{eq:integration_BS}) in Equation~(\ref{eq:alpha_ind}), we get the liifting line Equation:

\begin{equation}\label{eq:LL_equation}
	\Gamma(\vecl{y}) = \frac{1}{2}c(\vecl{y})C_{l\alpha}\Big[|\vecl{U}|\alpha(\vecl{y}) - \frac{1}{4\pi}\sum_{i}\Big(-d\Gamma_{i}\dfrac{(\vecl{y}-\vecl{y}_{i})\times\hat{\vecl{e}}_{x'_{w}}}{\left|\vecl{y}-\vecl{y}_{i}\right|^2}\Big) \cdot \hat{\vecl{e}}_{n}\Big]
\end{equation}
To satisfy the solenoidal property of the vorticity field, the circulation $\Gamma$ must form closed loops. This means that the circulation of the vortex tubes can be computed from the variations of $\Gamma$ along the lifting line. For a given vortex tube $i+1$ between the points $i$ and $i+1$ of the lifting line, the circulation of a shed tube $d\Gamma_{i+1}$ is equal to
\begin{equation}\label{eq:vortex_tube}
d\Gamma_{i+1} = \Gamma_{i+1}-\Gamma_i
\end{equation}
where $\Gamma_i$ is the local circulation at the $i^\text{th}$ point of the lifting line.\\
The circulation of the lifting line is computed iteratively. Starting from an initial guess, the induced velocities are computed at each point of the lifting line, with the contribution of each vortex tube from Biot-Savart law. The angle of attack is then modified with the new local flow conditions and new values are obtained from Equation~(\ref{eq:alpha_ind}). The circulation of the vortex tubes are then computed from Equation~(\ref{eq:vortex_tube}), thus closing the loop.\\
Once all the circulations are computed at every time step, the aerodynamic force acting on each discretized point of the lifting line computed as
\begin{equation}\label{eq:aero_force}
d\vecl{F}_i= \rho \Gamma_{i}(\vecl{U_{r}} \times \hat{\vecl{e}}_t)dl_{i}
\end{equation}
Finally, the global forces acting on the wing, can be evaluated in order to close the system of Equation~(\ref{eqn:eom2}), by summing each contribution and computing the corresponding pitching moment, i.e.
\begin{equation}\label{eq:aero_force}
\begin{aligned}
F_{x'} &= \sum_{i=1}^{N} (d\vecl{F}_i) \cdot \hat{\vecl{e}}_{x'}\\
F_{z'} &= \sum_{i=1}^{N} (d\vecl{F}_i) \cdot \hat{\vecl{e}}_{z'}\\
M_{y'} &= \sum_{i=1}^{N} (\vecl{GP_i})\times(d\vecl{F}_i)\cdot \hat{\vecl{e}}_{y'}
\end{aligned}
\end{equation}

\section{Flight dynamics analysis}\label{sec:methods}
We can now address how to solve Equations~(\ref{eqn:eom2}), with the aerodynamic terms that are computed according to Equation~(\ref{eq:aero_force}). Since it has been previously established that the wings follow a periodic trajectory governed by harmonic functions, the solution of Equation~(\ref{eqn:eom2}) must be a limit cycle, for some particular initial conditions. Therefore, studying the stability of these solutions requires a specific formalism, and we will use here Floquet stability analysis. This section investigates the key features of periodic orbits, presenting a numerical algorithm both for identifying limit cycles and quantifying their stability from the so-called corresponding Floquet multipliers.

\subsection{Stability of limit cycles}\label{sec:periodicstability}
The objective is to find a limit cycle corresponding to a trimmed configuration of the bird flight, i.e. the flight configuration corresponding to a periodic trajectory of all the state variables. Moreover, we are interested in assessing the stability of such a limit cycle.
Considering a generic non-autonomous system of ordinary differential equations in the form
\begin{equation}
\dot{\mathbf{x}} = \mathbf{v}(\mathbf{x}, t)
\label{eqn:odena}
\end{equation}
the limit cycle is a particular solution such that $\textbf{x}^*(t) = \textbf{x}^*(t+T)$ with $T$ being the cycle period. \\Limit cycles can be stable or unstable, depending on whether a perturbed initial value tends to be dynamically attracted or repelled by the periodic orbit.

\begin{figure}[htbp]
\centering
\includegraphics[width=.35\textwidth]{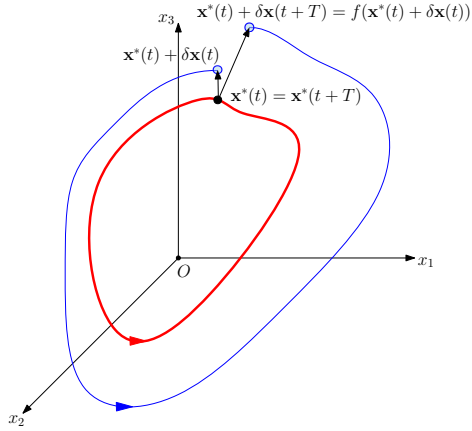}
\caption{Illustration of a periodic orbit (red) and a first integration over one period of the neighboring trajectory (blue).}
\label{fig:stabilityscheme}
\end{figure}
Their stability is assessed by Floquet theory~\cite{floquet1883}, and governed by the eigenvalues $\Lambda_{i}$ of the Jacobian matrix (or Monodromy matrix) $ \mathbb{J}(\mathbf{x}_0) \Big \rvert_{t_0}^{t_0 + T}$ that quantifies how a small perturbation out of the limit cycle is deformed by the flow, after a period $T$~\cite{ChaosBook, strogatz2018, seydel2009}. Calling $\mathbf{x}_0$ the initial condition, this Jacobian matrix is thus the result of the integration of the following system up to time $t=T$
\begin{equation}
\begin{aligned}
	\frac{d\mathbb{J}}{dt}(\mathbf{x}_0) \Big \rvert_{t_0}^{t}&= \mathbb{A}(\mathbf{x}, t) \mathbb{J}(\mathbf{x}_0) \Big \rvert_{t_0}^{t}\\
	\mathbb{J} (\mathbf{x}_0)\Big \rvert_{t_0}^{t_0} &= \mathbb{I}
\label{eqn:jacobiancomputation2}
\end{aligned}
\end{equation}
where the matrix 
\begin{equation}
\mathbb{A}(\mathbf{x}, t) = \frac{\partial}{\partial x_j}v_{i(\mathbf{x}, t)}
\label{eqn:stability_matrix}
\end{equation}
 is called the Stability Matrix~\cite{ChaosBook} and is $T$-periodic on the limit cycle.  These eigenvalues are also called \textit{Floquet Multipliers}. In other words, the Jacobian Matrix maps infinitesimal perturbations embedded within a sphere around a specific point of the limit cycle at a given time $(\mathbf{x_0}, t_0)$, to a stretched ellipsoid after a time $t$~\cite{ChaosBook}. This stretching ratio is governed by the Floquet Multipliers, and the stretching directions by its eigenvectors. 
Floquet multipliers have the property to be independent of the choice of $\mathbf{x}_0$ on the limit cycle, while the Floquet matrix and its eigenvectors depend on it~\cite{lust2001}.\\
If the system was autonomous, i.e. of the form $\mathbf{\dot{x}} = \mathbf{v}(\mathbf{x})$, one of the Floquet multipliers would systematically be equal to one and its eigenvector would be tangent to the limit cycle at $\mathbf{x}_0$. In the literature, this eigenvalue is often called the \textit{trivial} or \textit{marginal} multiplier, and a periodic solution is said asymptotically stable if all Floquet multipliers except this one are strictly smaller than one in absolute value. For a non-autonomous systems like the one considered here, all Floquet multipliers are non-trivial, and therefore stability requires that all Floquet multipliers to be smaller than one in absolute value~\cite{taha2012}. 


\subsection{Multiple-shooting method}
A multiple shooting algorithm is employed in order to identify the periodic orbits corresponding to trimmed flight, and to simultaneously compute their stability through their Floquet multipliers. \\We use a multiple-shooting scheme first proposed by~\cite{lust2001}, which was a modification of~\cite{keller1968}. This algorithm is adapted to our case with the advantage that the limit cycle period is known, since it must be equal to the flapping one~\cite{dietl2008}.  
\begin{figure}[htbp]
\centering
\includegraphics[width=.5\textwidth]{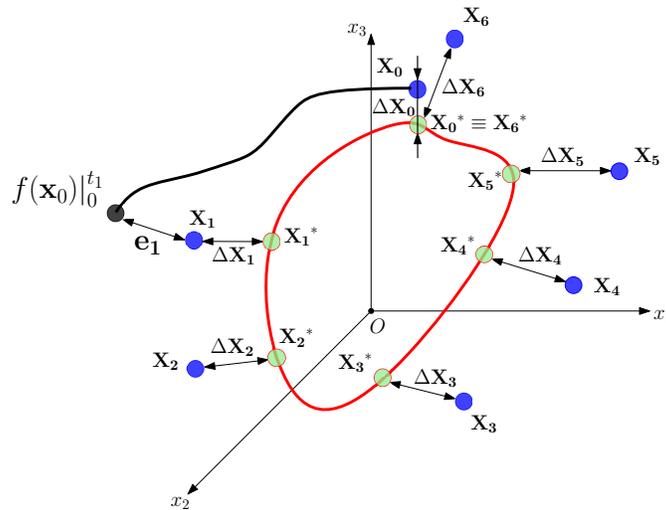}
\caption{Generic computational scheme describing the multiple-shooting method of a phase-space $\mathcal{M} \in  \rm I\!R^{3}$ (without loss of generalities). The black trajectory is traced out by integrating the equations from the guessed points (blue dots). The asymptotic limit cycle is represented in red, and the points belonging to it in green.  }
\label{fig:multipleshooting}
\end{figure}\\
Considering a generic $T$-periodic non-autonomous system described by Equation~(\ref{eqn:odena}), this multiple-shooting method splits the limit cycle into several points and computes the time integration from one point to the following along the trajectory, as illustrated in Figure~\ref{fig:multipleshooting}.\\
By time-integration of the system dynamics, the point $\mathbf{x}^*_{i+1} $ is mapped to the point  $\mathbf{x}^*_{i} $ by
\begin{equation}
\mathbf{x}^*_{i+1} = f(\mathbf{x}^*_i)  \big \rvert_{t_{i}}^{t_{i}+\tau} = f(\mathbf{x}_i + \Delta\mathbf{x}_i) \big \rvert_{t_{i}}^{t_{i}+\tau}
\label{eqn:multishooting2}
\end{equation}
Computing the  Taylor first order expansion of the right hand side of Equation~(\ref{eqn:multishooting2}), the point $\mathbf{x}^*_{i+1}$ can be expressed as function of the guessed points only
 \begin{equation}
\mathbf{x}_{i+1} + \Delta\mathbf{x}_{i+1}  =f(\mathbf{x}_i) \big \rvert_{t_{i}}^{t_{i}+\tau} + \mathbb{J} (\mathbf{x}_i) \Big \rvert_{t_{i}}^{t_{i}+\tau}\cdot\Delta\mathbf{x}_i
\label{eqn:multishooting3}
\end{equation}
where $\mathbb{J} \big \rvert_{t_{i}}^{t_{i}+\tau}(\mathbf{x}_i)$ is the Jacobian matrix defined in Equation~(\ref{eqn:jacobiancomputation2}).
Re-arranging Equation~(\ref{eqn:multishooting3}) as
 \begin{equation}
	 \mathbb{J}(\mathbf{x}_i) \Big \rvert_{t_{i}}^{t_{i}+\tau}\cdot\Delta\mathbf{x}_i -\Delta\mathbf{x}_{i+1} = \underbrace{-\big(f(\mathbf{x}_i)\big \rvert_{t_{i}}^{t_{i}+\tau} - \mathbf{x}_{i+1}\big)}_{Error}
\label{eqn:multishooting4}
\end{equation}
the multiple-shooting scheme can be derived
\begin{equation}
\underbrace{
\begin{pmatrix}
\mathbb{J} (\mathbf{x}_0) \Big \rvert_{0}^{\tau}  & - \mathbb{I}& 0& \dots& 0 \\
\\ 
0 & \mathbb{J} (\mathbf{x}_1)\Big \rvert_{t_{1}}^{t_{1}+\tau}& - \mathbb{I}  & \dots & 0\\
\vdots & \vdots & \ddots & \ddots & \vdots \\
0 & 0 &\dots & \mathbb{J}(\mathbf{x}_{m-1})\Big \rvert_{t_{m-1}}^{T}  & - \mathbb{I}\\
- \mathbb{I} & 0 &\dots & 0 &  \mathbb{I}\\
\end{pmatrix}}_{\mathbf{M}\ [n \times M, n \times M]}
\underbrace{
\begin{pmatrix}
\Delta \mathbf{x}_{0}\\
\Delta \mathbf{x}_{1}\\
\vdots\\
\vdots\\
\vdots\\
\Delta \mathbf{x}_{m-1}\\
\Delta \mathbf{x}_{m}
\end{pmatrix}}_{\Delta\mathbf{x}\ [n \times M]}=
\underbrace{-\begin{pmatrix}[2]
f(\mathbf{x}_0) \big \rvert_{0}^{\tau}- \mathbf{x}_1 \\
 f(\mathbf{x}_1) \big \rvert_{t_{1}}^{t_{1}+\tau}- \mathbf{x}_2 \\
\vdots\\
(\mathbf{x}_{m-1}) \big \rvert_{t_{m-1}}^{T} - \mathbf{x}_m\\
\mathbf{x}_{m}- \mathbf{x}_0\\
\end{pmatrix}}_{\mathbf{E}\ [n \times M]}
\label{eqn:shootingscheme}
\end{equation}
Calling $n$ the number of states of the dynamical system and $M$ the amount of points employed in the multiple-shooting, $\mathbf{M}$ is the Multiple-shooting matrix of dimension $[n \times M, n \times M]$, $\Delta\mathbf{x}$ the unknown vector of dimension $[n \times M]$ and $\mathbf{E}$ the error vector of dimension $[n \times M]$.\\
By expressing Equation~(\ref{eqn:shootingscheme}), in a compact form 
\begin{equation}
\mathbf{M}(\mathbf{x}_i) \mathbf{\Delta \mathbf{x}} = \mathbf{E}(\mathbf{x}_i)
\label{eqn:multishootingcompact}
\end{equation}
Finding the solution of Equation~(\ref{eqn:multishootingcompact}), consists in finding $\mathbf{x}_i^* \in \rm I\!R^{n}$ such that $\mathbf{E}(\mathbf{x}_i^*) = \mathbf{0}$ and this can be solved with an iterative scheme. 
A Newton's method  can be derived to iterate on Equation~(\ref{eqn:multishootingcompact}) until convergence to 0. However, the main drawback of this method, is the high sensitivity on the choice of the initial guess $\mathbf{x}^{(0)}_i$. It can be mathematically proved that Newton's method quadratically converges only if the choice of the initial conditions is sufficiently close to the solution (for a proof of this Theorem, please refer to~\cite{quarteroni2010}, chap. 7).\\
Consequently, we implemented a modified Newton's method, in order to improve the robustness of the scheme with respect to the initial value. The proposed method relies on the Levenberg–Marquardt algorithm~\cite{marquardt1963} (LMA). Such an implementation of LMA in a multi-shooting code was already adopted by Dednam and Botha~\cite{dednam2015} and the code was validated with both autonomous and non-autonomous systems. \\
The estimation of the unknown vector $\delta \mathbf{x}$, used to update the state variables at the generic iteration step $k$, is computed as follows
\begin{equation}
\left[ \mathbf{M}^{T}\mathbf{M} + \lambda \text{diag}(\mathbf{M}^{T}\mathbf{M}) \right]\mathbf{\delta \mathbf{x}} =  \mathbf{M}^{T} \mathbf{E}
\label{eqn:lma}
\end{equation}
where $\lambda$ is a non-negative, adaptive damping parameter, and a candidate algorithm is presented in Algorithm~(\ref{alg:lma}).
\begin{algorithm}
\caption{Levenberg–Marquardt}\label{euclid}
\begin{algorithmic}[1]
\State $\mathbf{x}^{(0)}_i \gets \vars{guessed\_points}$
\State $k  \gets 0$
\Function{LMA}{$\vars{guessed\_points}, \vars{tolerance}$}
\State $\lambda  \gets \textit{set}$
\State $\mathbf{M}(\mathbf{x}^{(o)}_i), \mathbf{E}(\mathbf{x}^{(o)}_i) \gets \textit{Compute}$
\While {$|\mathbf{E}(\mathbf{x}^{(k)}_i)|> \epsilon$} 
\State $\mathbf{M}(\mathbf{x}^{(k)}_i), \mathbf{E}(\mathbf{x}^{(k)}_i) \gets \textit{Compute}$
\State  $\left[ \mathbf{M}^{T}\mathbf{M} + \lambda \text{diag}(\mathbf{M}^{T}\mathbf{M}) \right]\mathbf{\delta \mathbf{x}} =  \mathbf{M}^{T} \mathbf{E} \gets \textit{Solve}$
\State $\mathbf{x}^{(k+1)} \gets \mathbf{x}^{(k)} + \delta \mathbf{x}^{(k)}$
\State $\mathbf{E}(\mathbf{x}^{(k+1)}_i) \gets \textit{Compute}$
\If{$\min{ | \mathbf{E}(\mathbf{x}^{(k+1)}_i)|} < \min{|\mathbf{E}(\mathbf{x}^{(k)}_i)| }$}
\State $\lambda = \lambda/ \nu$
\Else
\State $\lambda = \lambda* \nu$
\EndIf
\EndWhile
\State
\Return ${\mathbf{x}^{(k+1)}_i}$
\EndFunction
\end{algorithmic}
\label{alg:lma}
\end{algorithm}
When the trajectory eventually converges to the limit cycle, the Jacobian matrix of the whole limit cycle obeys the semigroup property and can be expressed as the product of the submatrices of Equation~(\ref{eqn:shootingscheme}), i.e.
\begin{equation}
\mathbb{J} (\mathbf{x}_0) \Big \rvert_{0}^{T} =    \mathbb{J} (\mathbf{x}_{m-1}) \Big \rvert_{t_{m-1}}^{T} \cdots  \mathbb{J} (\mathbf{x}_1) \Big \rvert_{t_{1}}^{t_{1}+\tau}\cdot \mathbb{J} (\mathbf{x}_0) \Big \rvert_{0}^{\tau}
\label{eqn:semigroup}
\end{equation}
\subsection{Computation of the Jacobian matrix}\label{subsec:jacobiancomputation}
Two concurrent approaches can be used to evaluate the Jacobian matrix and build the diagonal blocks of the multiple-shooting matrix $\mathbf{M}$ in Equation~(\ref{eqn:shootingscheme}): the first one relies on a semi-analytical approach, while the second one relies on numerical computations only. Both methods are implemented in our code, and a comparison of the results is reported in Appendix~\ref{appendix:jacobian}. 
The semi-analytical approach is the one we used in the following simulations, and it is here described in details. The Jacobian matrix is obtained by solving the variational Equation~(\ref{eqn:jacobiancomputation2}) and the state Equation~(\ref{eqn:odena}), i.e.
\begin{equation}
   \begin{pmatrix} \dot{\mathbf{x}}\\ \dot{\mathbb{J}} \end{pmatrix} = \begin{pmatrix} \mathbf{v}(\mathbf{x}, t)\\ \mathbb{A}(\mathbf{x}, t) \ \mathbb{J} \end{pmatrix}
\label{eqn:jacobianreshaped_formula}
\end{equation}
with the initial condition
\begin{equation}
   \begin{pmatrix} {\mathbf{x}(t_0)}\\ {\mathbb{J}^{0}} \end{pmatrix} = \begin{pmatrix} \mathbf{x}_0 \\ \mathbb{I} \end{pmatrix}
\label{eqn:initialconditionjacobian}
\end{equation}
This approach requires to simultaneously solve $(n+n^{2})$ ordinary differential equations~\cite{seydel2009}. The solution of this system corresponds to the Jacobian matrix of a generic trajectory at time $t_f$, obtained from an initial condition at time $t_0$.\\
In our particular case, the matrix $\mathbb{A}(t)$ accordingly to Equation~(\ref{eqn:stability_matrix}) is
\begin{equation}
\mathbb{A}(\mathbf{x}(t), t) = 
\begin{pmatrix}
\frac{1}{m}\frac{\partial{F_{x'}}}{\partial{u}}  & \frac{1}{m}\frac{\partial{F_{x'}}}{\partial{w}} - q& - w + \frac{1}{m}\frac{\partial{F_{x'}}}{\partial{q}}& -g\cos{\theta} + \frac{1}{m}\frac{\partial{F_{x'}}}{\partial{\theta}}\\
q + \frac{1}{m}\frac{\partial{F_{z'}}}{\partial{u}}  & \frac{1}{m}\frac{\partial{F_{z'}}}{\partial{w}}&  u + \frac{1}{m}\frac{\partial{F_{z'}}}{\partial{q}} & g\sin{\theta} + \frac{1}{m}\frac{\partial{F_{z'}}}{\partial{\theta}}\\
\frac{1}{I_{yy}}\frac{\partial{M_{y'}}}{\partial{u}}  & \frac{1}{I_{yy}}\frac{\partial{M_{y'}}}{\partial{w}}& \frac{1}{I_{yy}}\frac{\partial{M_{y'}}}{\partial{q}}  & \frac{1}{I_{yy}}\frac{\partial{M_{y'}}}{\partial{\theta}}\\
0 & 0 &1 & 0 \\
\end{pmatrix}
\label{eqn:stabmatrixeom}
\end{equation}

Given the complexity of our coupled system of Equation~(\ref{eqn:eom2}), the main disadvantage of using this approach is the need of computing the derivatives of the aerodynamic forces.
In other words, even by using such an analytical approach, a numerical differentiation is necessary to compute these derivatives in order to solve Equation~(\ref{eqn:jacobianreshaped_formula}).\\
\subsection{Numerical parameters and wingtip trajectory}
\begin{figure}[!htbp]
\begin{center}
\includegraphics[scale=.3]{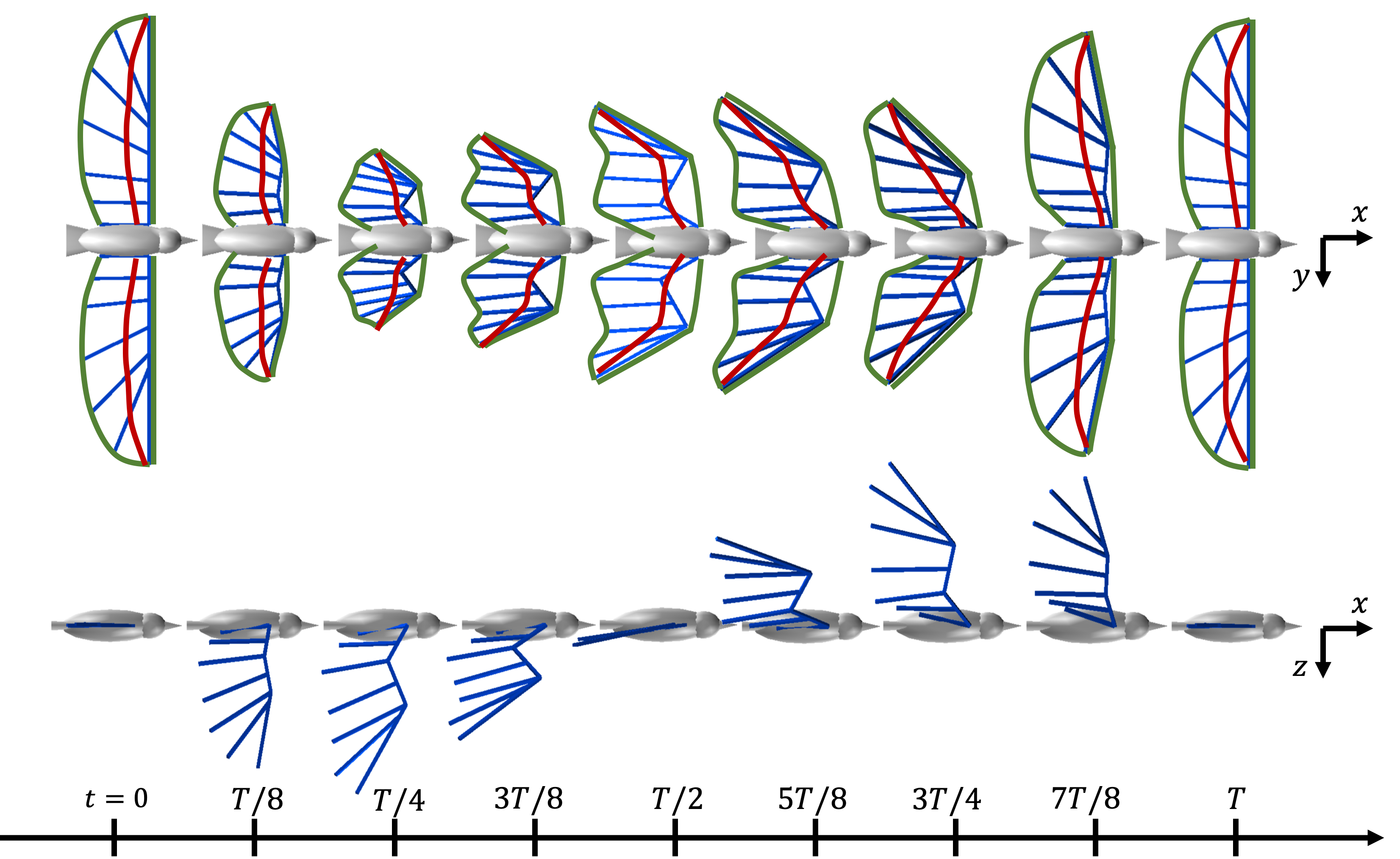}
\caption{Snapshots of the wing kinematics over a flapping cycle. The three rigid segments and the seven "master-feathers" are represented in blue. The green lines represent the leading and trailing edges and the red line captures the lifting line where the wing aerodynamic forces are computed.}
\label{fig:wing_kine_fig}
\end{center}
\end{figure}

The aerodynamic model can be adapted to large scale flapping fliers, and we implemented lengths of the bones and feathers to match those of the northern bald ibis (\emph{Geronticus eremita}). This particular bird has been chosen because it has a high aspect ratio wing -- which is well suited for the lifting line approach used -- and uses non-stop flapping flight.\\
The parameters governing the wing kinematics described by Equation~(\ref{eqn:general_joint}) are constrained to follow the wing kinematics of real birds. No accurate data about the kinematics of ibises and other large birds are available in the literature. We thus re-scale the kinematics of a pigeon, reported in~\cite{tobalske1996}. The typical wingbeat frequency is retrieved from~\cite{vanderberg1995} and we tuned the wingbeat amplitude accordingly, in order to keep the angle of attack in a realistic range.\\ 
Based on the aforementioned observations, all simulations reported in the rest of this paper have been computed with numerical parameters gathered in Table~\ref{tab:parameters}. The wingbeat frequency is taken equal to $f=4Hz$.
\begin{table}[!http]
\centering
\subtable [Parameters describing the bird morphology.\label{tab:morphology}]{
\begin{tabular}{|l|c|}
    \hline
    \multicolumn{2}{|c|}{\textbf{Bird body}}\\ \hline 
    Mass $(m_{b})$, $[kg]$											&				1.2\\
    Moment of Inertia  $(I_{y})$, $[kg \cdot m^{2}]$&0.1\\ \hline
    \multicolumn{2}{|c|}{\textbf{Bird wing}}\\ \hline 
        Wingspan $(b)$, $[m]$						&				1.35\\
     Mean aerodynamic chord $(\overline{c})$, $[m]$		&	0.2\\ 
	Arm bone length $(l_a)$, $[m]$				& 				0.134\\
	Forearm bone length $(l_f)$, $[m]$		& 				0.162\\
	Hand bone length $(l_h)$, $[m]$			& 				0.084\\ \hline
	\multicolumn{2}{|c|}{\textbf{Bird feathers}}\\ \hline 
	Primary feather 1 $(l_{p1})$, 			$[m]$			&				0.25\\
	Primary feather 2 $(l_{p2})$,		    $[m]$			&				0.275\\
	Primary feather 3 $(l_{p3})$,		    $[m]$			&				0.25\\
	Secondary feather 1 $(l_{s1})$,		$[m]$			&				0.225\\
	Secondary feather 2 $(l_{s2})$,		$[m]$			&				0.2\\
	Secondary feather 3 $(l_{s3})$,		$[m]$			&				0.175\\
	Secondary feather 4 $(l_{s4})$,		$[m]$			&				0.15\\ 

    \hline
\end{tabular}
}\qquad\qquad
	\subtable[Kinematic parameters governing the wing motion, defined according to Equation~(\ref{eqn:general_joint}). Note that the amplitude of the shoulder joint along the $x$ axis, denoted $A_{s,x}$ in the table, varies across the experimental conditions in order to guarantee trimmed flight.\label{tab:kinematics}]{
\begin{tabular}{|l|c|c|c|}
\hline
\textbf{Joint} & 		$q_{0} [\deg]$&             $A[\deg]$&                     $\phi[\deg]$  \\
\hline
{Shoulder} $y$ & 		$11.5$&             $0.8$&                     $-90$  \\
{Shoulder} $x$ & 		$0$&             $A_{s,x}$&                     $180$  \\
{Shoulder} $z$ & 		$19$&             $20$&                     $90$  \\
\hline
{Elbow} $x$ & 		$30$&             $30$&                     $-90$  \\
\hline
{wrist} $z$ & 		$-30$&             $30$&                     $90$  \\
{wrist} $y$ & 		0&             $30$&                     $-90$  \\
\hline
\end{tabular}
}
\caption{Numerical parameters used in the simulations.}
\label{tab:parameters}
\end{table}\\
The resulting wing kinematics are also pictured in Figure \ref{fig:wing_kine_fig}, with a shoulder amplitude equal to $A_{s,x}=42\deg$. 

\begin{figure}[!htbp]
\centering
\includegraphics[width=.5\textwidth]{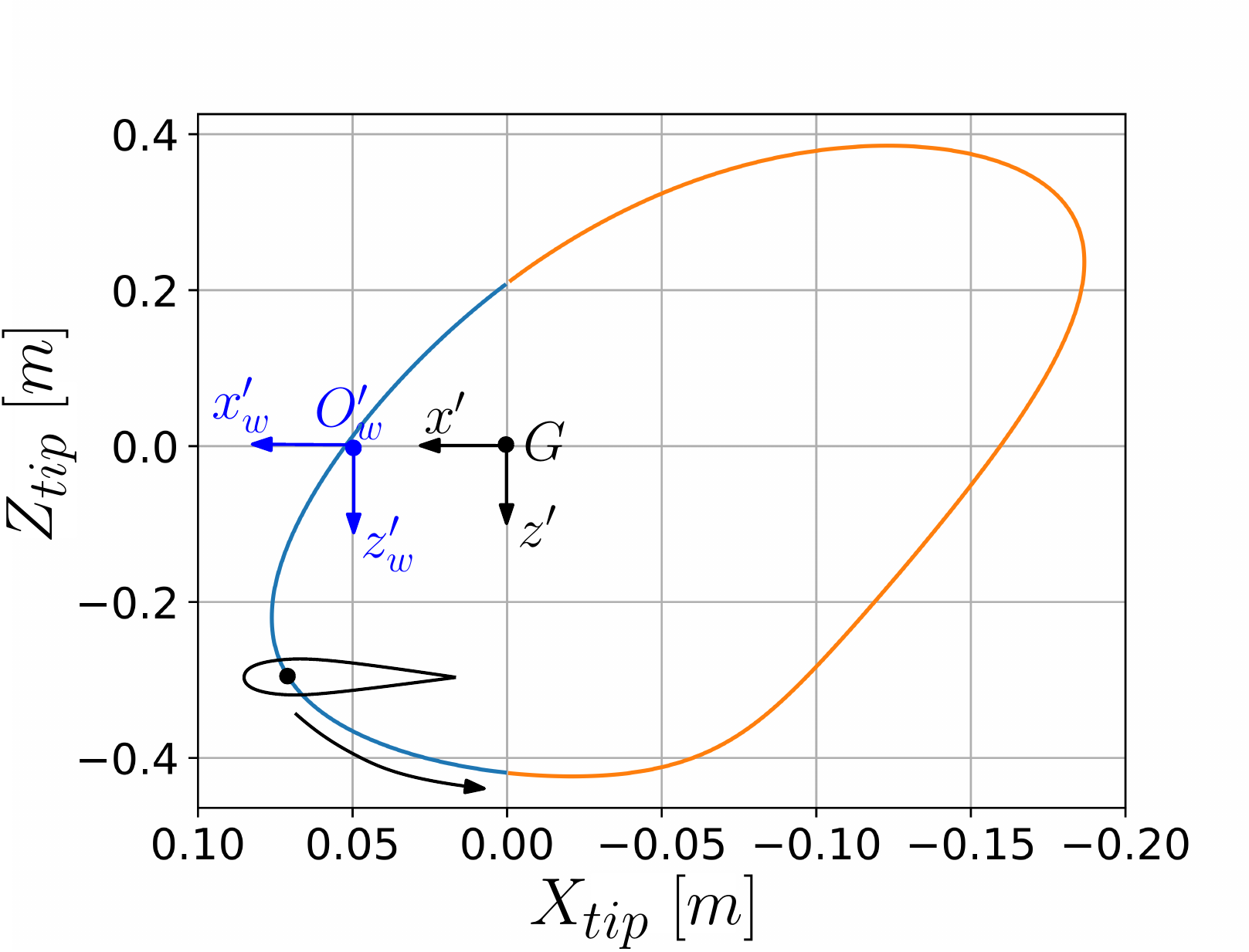}
\caption{Tip path trajectory over a wing beat cycle. Blue line: position of the lift ahead of the center of mass, contributing for a nose up (positive) pitching moment; orange line: position of the lift ahead of the center of mass, contributing for a nose down (negative) pitching moment.}
\label{fig:tip_trajectory}
\end{figure}

The reference position of the wing-bone frame is vertically aligned with the center of mass, and placed at a certain distance ahead of it. This default distance is 5cm, although a dedicated sensitivity analysis on this parameter is conducted later on (see Section~\ref{results}). Using the kinematic parameters of Table~\ref{tab:parameters} in Equation~(\ref{eqn:general_joint}) for each joint, and considering the relative position between the wing-bone frame and the body frame, the resulting tip trajectory is shown in Figure~\ref{fig:tip_trajectory}. Considering a counter-clockwise motion, the blue trajectory represents the region where the lifting line lies ahead of the center of mass $G$. Assuming a positive lift for the whole wingbeat period, this region provides a positive pitching moment due to the lift (nose-up). The tip positions corresponding to the orange segment of the trajectory correspond to the region where the lifting line lies behind the center of mass, generating a negative pitching moment due to the lift effect, under the aforementioned assumption. This is an important feature of our model since no additional lifting surface such as a tail is considered in the present study. Consequently, a necessary condition for the existence of a limit cycle is that the wing itself generates both positive and negative pitching moment, in order to achieve rotational equilibrium of the bird body over the cycle.

The numerical parameters employed in multiple-shooting algorithm are reported in Table~\ref{tab:shootingset} and the numerical integration has been performed using a 4th order Runge-Kutta scheme.
\begin{table}[htbp]
\centering
\begin{tabular}{|l|c|}
\hline
\multicolumn{2}{|c|}{\textbf{Multiple-Shooting settings}}\\ \hline 
{ Integrator order}  				& 		4 \\
{Time steps over a period} 	& 		150 \\
Amount of points ($M$) 		& 		5\\
{Iteration error}  & 	$1e-5$\\
\hline
\end{tabular}
\caption{Numerical parameters used in the Multiple shooting algorithm.}
\label{tab:shootingset}
\end{table}\\

\section{Results: identification, stability, and sensitivity analysis of limit cycles}\label{results}
This section reports three experiments that were conducted to validate the capacity of the developed framework to identify limit cycles, assess their stability, and quantify the sensitivity of flight regime and stability with respect to kinematic and morphological parameters 

\subsection{Experiment 1: representative limit cycle and stability analysis}
\begin{figure}[!htbp]
\begin{center}
\begin{minipage}[c]{.45\textwidth}
\centering
\includegraphics[width=1.\textwidth]{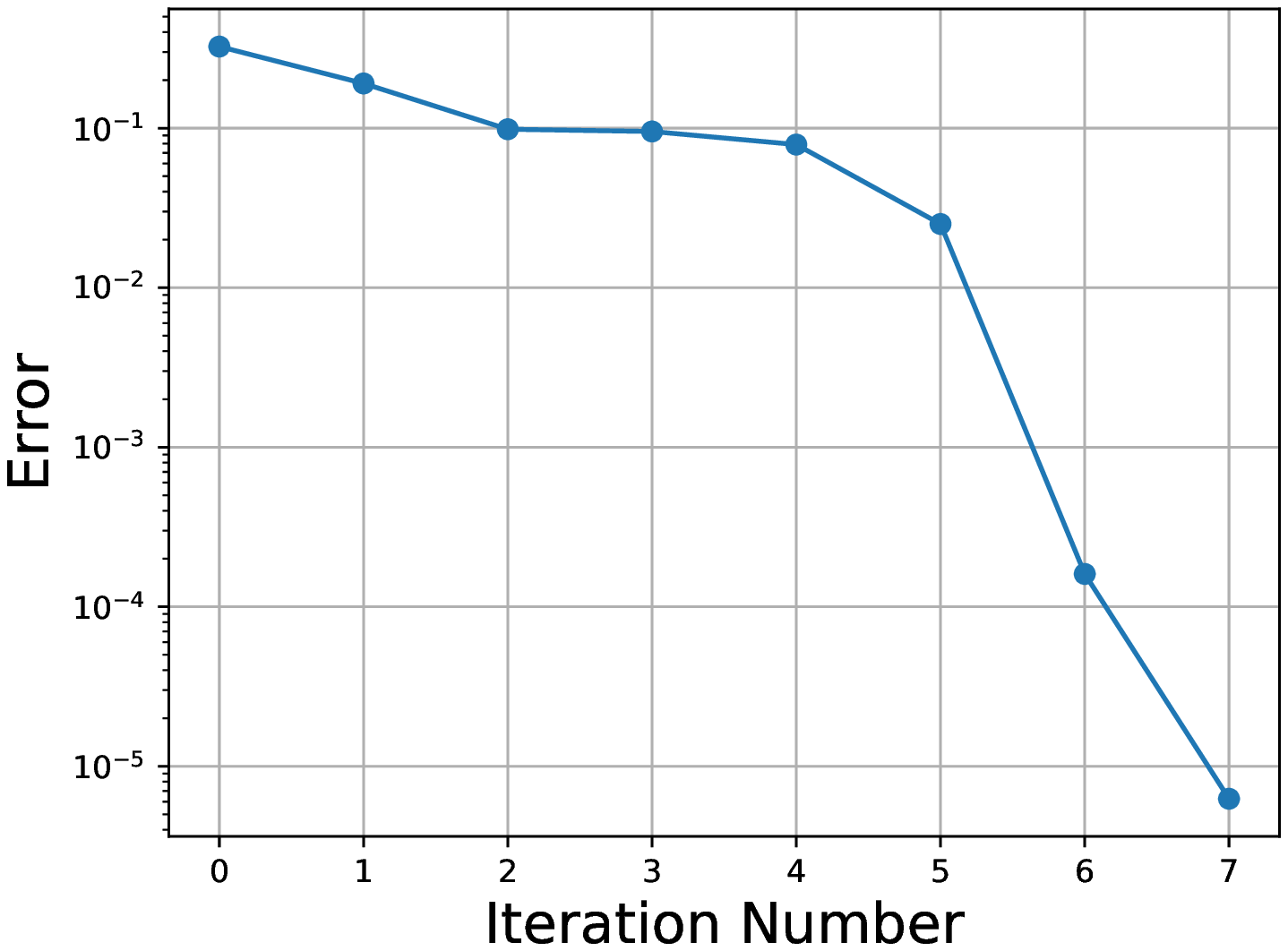}
\end{minipage}%
\hspace{10mm}%
\begin{minipage}[c]{.45\textwidth}
\centering
\includegraphics[width=1.\textwidth]{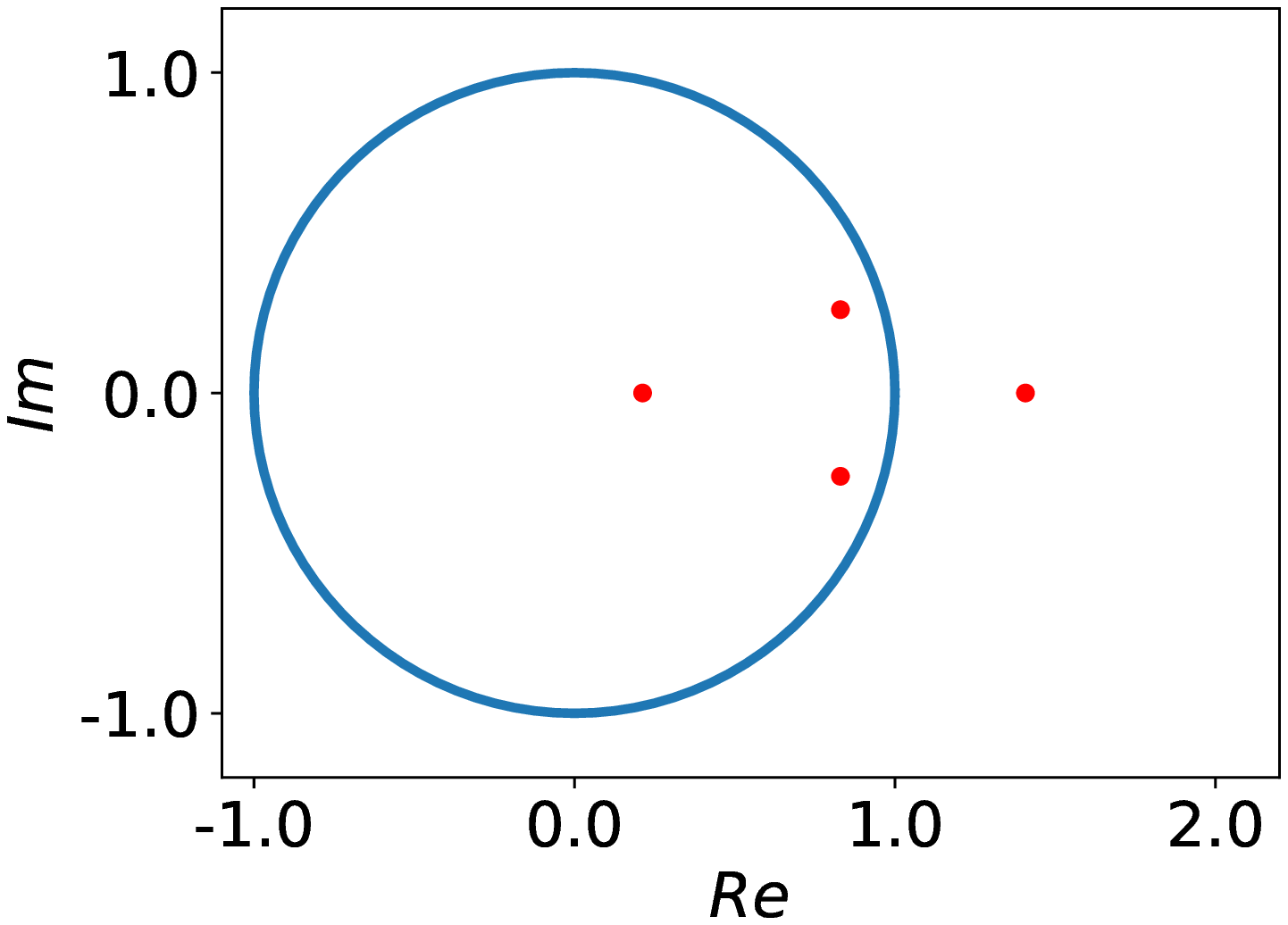}
\end{minipage}
\end{center}
\caption{Convergence error (left) and Floquet multipliers (right) for the reference case with the imposed kinematics of Table~\ref{tab:parameters}.}
\label{fig:sim_results1}
\end{figure}

A representative limit cycle solution is reported here, as the result of a multiple-shooting computation. This solution corresponds to the reference case in which the kinematics is described by the governing parameters of Table~\ref{tab:parameters}, with $A_{s,x}=42\deg$.
Convergence analysis and consequently stability results are shown in Figure~\ref{fig:sim_results1}. In particular, rapid convergence is obtained, after 7 iterations. The error is evaluated as the $\max\{{|\mathbf{E}(\mathbf{x})|}\}$ of Equation~(\ref{eqn:multishootingcompact}). Such configuration presents one expanding eigenvalue, which leads the system to be unstable. 
The numerical values of the Floquet multipliers, are also reported in Table~\ref{tab:eigenvalues}
\begin{table}[!htbp]
\centering
\begin{tabular}{|l|c|c|}
\hline
Multiplier		 & 		Value \\
\hline
$\Lambda_1 $ & 		$1.40+0j$ \\
$\Lambda_2 $ & 		$0.829+0.260j$\\
$\Lambda_3 $ & 		$0.829-0.260j$\\
$\Lambda_4 $ & 		$0.212 + 0j$\\
\hline
\end{tabular}
\caption{Floquet multipliers obtained for the reference case with the imposed kinematics of Table~\ref{tab:parameters}.}
\label{tab:eigenvalues}
\end{table}\\

The expanding eigenvalue has an absolute value of $\Lambda_{1}= 1.40$.  The related eigenvector is $e_{1} = [-0.69, 0.68, 0.12, 0.16]^{T} $ and therefore excites the perturbation along each eigenbase directions with the same order of magnitude.
Considering a flapping period of $T=0.25s$, the expanding Floquet exponent is$$\lambda_{1}=\frac{1}{T}\ln(\Lambda_{1})=1.34s^{-1}$$ and therefore the time needed for the perturbation to double its value is approximately $$t_{doubling} = \frac{\ln(2)}{\lambda_{1}}\approx 0.51s$$

\begin{figure}[!htbp]
\centering%
\includegraphics[width=1.\textwidth]{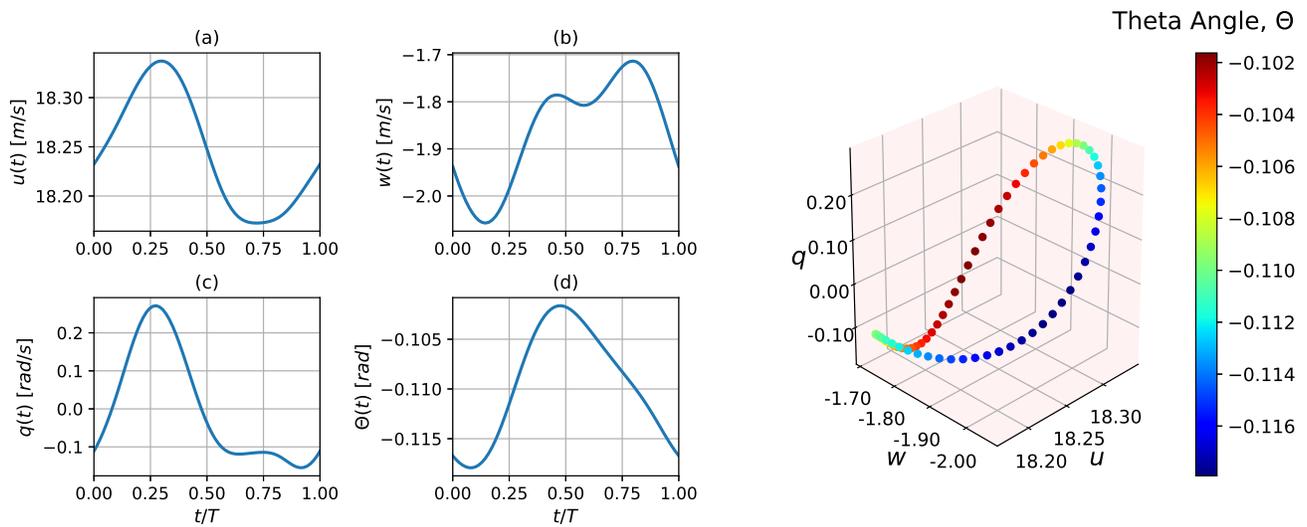}
\caption{Reference limit cycle solution. Trajectory of each state variable (left) and phase portrait (right). }
\label{fig:SVLimitCycle}
\end{figure}

\begin{figure}[!htbp]
\centering%
{\includegraphics[width=.8\textwidth]{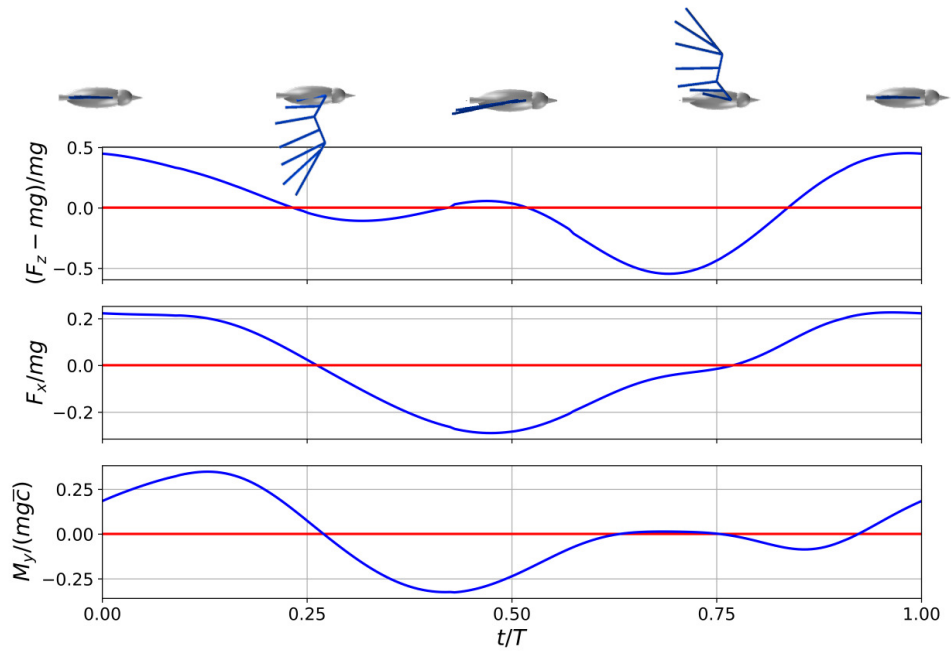}}
\caption{Dimensionless forces and pitching moment developed by the flier, expressed in the fixed frame $O(X,Z)$.}
\label{fig:aeroforces}
\end{figure}

This corresponds to approximatively two flapping period is thus larger than the one reported in previous studies focusing on smaller scale animals~\cite{dietl2008, Taylor2006}. Figure~\ref{fig:SVLimitCycle} pictures the periodic solution of the state variables describing this limit cycle in the phase space. This solution is a trimmed state of the bird, for the prescribed kinematics and morphology. The states are plotted with respect to one cycle in the moving body frame. At time zero the wing position corresponds to the middle of the downstroke.\\ 
On the right side of Figure~\ref{fig:SVLimitCycle} is pictured the 4-state limit cycle in the phase portrait. The periodicity in $\theta$ is plotted as a color map on the trajectory described by the three others state space variables.
\\
The aerodynamic forces and moment of the limit cycle solution are plotted in Figure~\ref{fig:aeroforces} and normalized with respect to $m_{b}g$ (forces) and $m_{b}g\overline{c}$ (moment).  The global action of the forces and pitching moment over one period is zero, confirming the state of trimmed flight, and limit cycle condition (zero acceleration over one period).
\subsection{Experiment 2: sensitivity analysis of the shoulder amplitude}
\subsubsection{Methods}
The multiple-shooting method has been applied to address the question of the gait sensitivity, since the framework can handle the analysis of several gait configurations. In particular, we exploited this to achieve a specific limit cycle solution corresponding to level flight. Indeed, the solution reported in Experiment 1 corresponds to trimmed flight, but not necessarily to level flight: trimmed flight might correspond to a flight regime with a non-zero averaged vertical velocity.\\
In order to achieve level flight, the mean vertical velocity with respect to the fixed frame has to be zero over the flapping period. Considering Figure~\ref{fig:birdframe}, the velocity components in the fixed frame are
\begin{equation}
\begin{aligned}
	U_{ff} &= \dot{X}  = u\cos{\theta} + w\sin{\theta}\\
	W_{ff}  &= -\dot{Z} = u\sin{\theta} - w\cos{\theta}
\label{eqn:fixedframeeqn}
\end{aligned}
\end{equation}
In particular, we report here a sensitivity analysis of the flapping gait as a function of one of the most important kinematic parameters, namely the wingbeat amplitude of the shoulder $A_{s,x}$.  Consequently, seeking for a level-flight configuration reduces to a single parametric study consisting in finding the shoulder amplitude $A^{*}_{s,x}$ that corresponds to a limit cycle whose mean vertical velocity is equal to zero, i.e.
\begin{equation}
\overline{W}_{ff}(A^{*}_{s,x})  = 0
\label{eqn:levelflight}
\end{equation}
Since ${W}_{ff}$ is a non-linear function of $A_{s,x}$, we rely on a Newton-Raphson method to find its root.
Finally, the climbing or descending ratio is identified by the trajectory angle, defined as $$\beta = \tan^{-1}{\frac{\overline{W}_{ff}}{\overline{U}_{ff}}}$$
\subsubsection{Results}
\begin{figure}[!htbp]
\centering
\includegraphics[width=.8\textwidth]{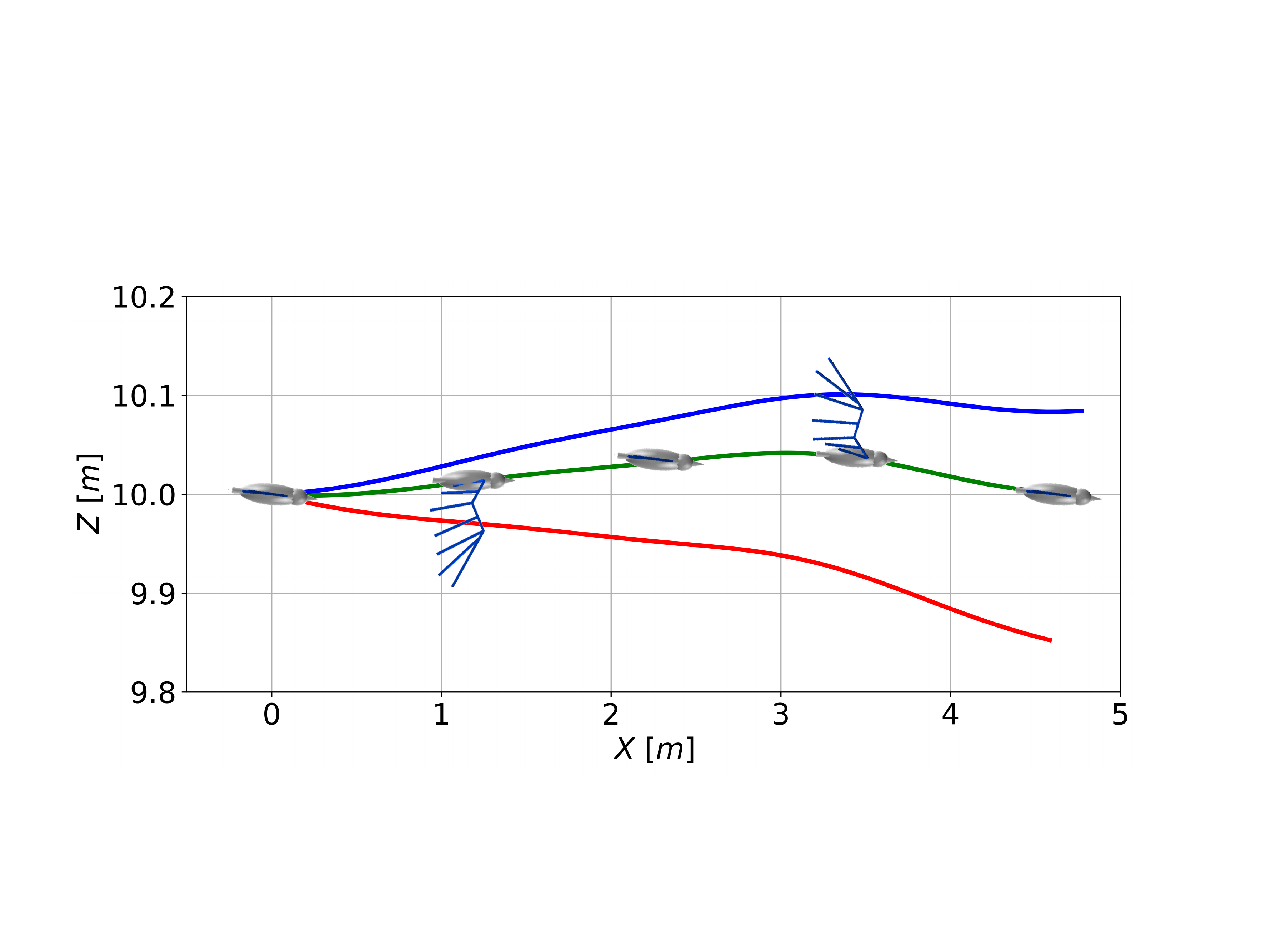}
\caption{Different flight trajectories over one flapping period. Red path: descending behavior with a shoulder amplitude of $42\deg$; green path: level flight solution obtained with a shoulder amplitude of $43.47\deg$; blue path: climbing behavior for a wingbeat amplitude of $44\deg$.}
\label{fig:trajectory_xz}
\end{figure}

\begin{figure}[!htbp]
\centering
\includegraphics[width=1.\textwidth]{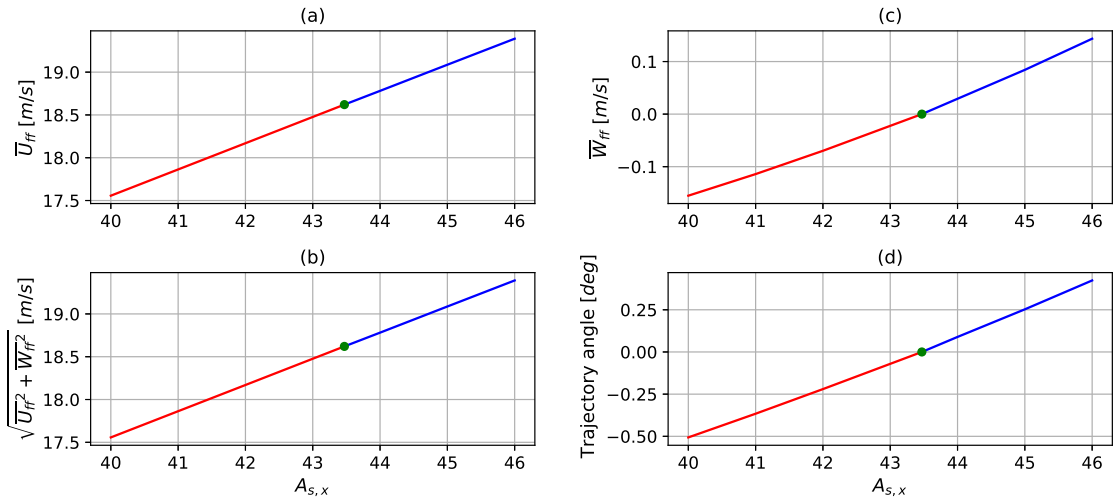}
	\caption{(a) Averaged horizontal veocity, (b) averaged vertical velocity, (c) averaged norm of the velocities, (d) trajectory angle.}
\label{fig:vel_amp}
\end{figure}

By applying the frame transformation of Equation~(\ref{eqn:fixedframeeqn}) and subsequently integrating the velocities, three different trajectories of the bird corresponding to three different flight conditions, are illustrated in Figure~\ref{fig:trajectory_xz}. 
The relationships between the shoulder amplitude, and the corresponding flight velocities and trajectory angles have has also been investigated for a interval of $[40\deg< A_{s,x}<46\deg]$, and results are plotted in Figure~\ref{fig:vel_amp}. A quasi-linear relationship is found with high sensitivity response of the amplitude parameter to the flight condition.\\Descending trajectories are achieved for amplitudes smaller that $43.47\deg$ corresponding to a negative vertical velocity, while climbing trajectories correspond to higher values of amplitudes. Also concerning the forward flight velocity, an increase in amplitude determines a linear increment of the flight speed, suggesting an active role of this parameter on the production of the thrust.
\begin{figure}[!htbp]
\centering
\includegraphics[width=.9\textwidth]{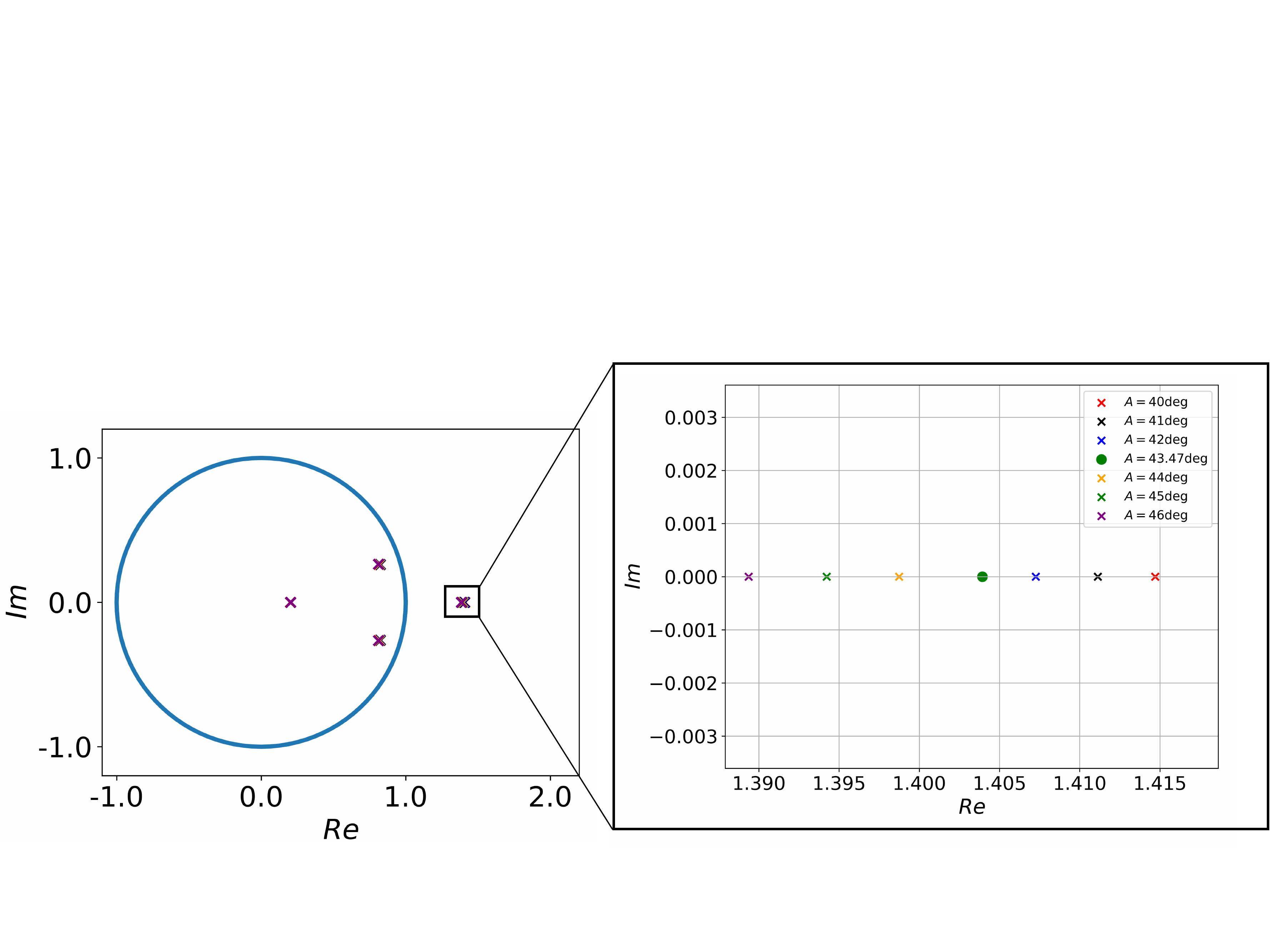}
\caption{Eigenvalues corresponding to several shoulder amplitudes and zoom on the unstable branch.}
\label{fig:eigen_amp}
\end{figure}
Finally we investigated the stability behavior for all the cases reported here. The Floquet multipliers are pictured in Figure~\ref{fig:eigen_amp} and this shows that the four multipliers are marginally affected by changing the shoulder amplitude.\\

\subsection{Experiment 3: Sensitivity analysis on the wing insertion point}
\subsubsection{Methods}
We now conduct a sensitivity analysis of stability as a function of the relative position between the center of mass and the insertion point of the wing in the body frame. This wing instertion point coincides with the rotational shoulder joint. Since the wing is free to sweep around the body center of mass, we need to revisit the standard definition of stability margin from the literature in flight mechanics. Considering Figure~\ref{fig:liftinglinemodel}, we redefined the stability margin $SM$ as position of the center of mass with respect to the wing root projection along $x'$, normalized by the mean aerodynamic chord, i.e.
\begin{equation}
SM = \frac{(G - O'_{w})}{\overline{c}}
\label{eqn:stab_margin}
\end{equation}

When $SM = 0\%$, the center of mass longitudinally coincides with the origin of the wing frame.
The interval of variation of the stability margin, has been explored in the range $-32.5 \%< SM< -17.5\%$.
\subsubsection{Results}

\begin{figure}[htbp]
\centering
\includegraphics[clip, trim=0.cm 1.cm 0cm 1.5cm, width=.7\textwidth]{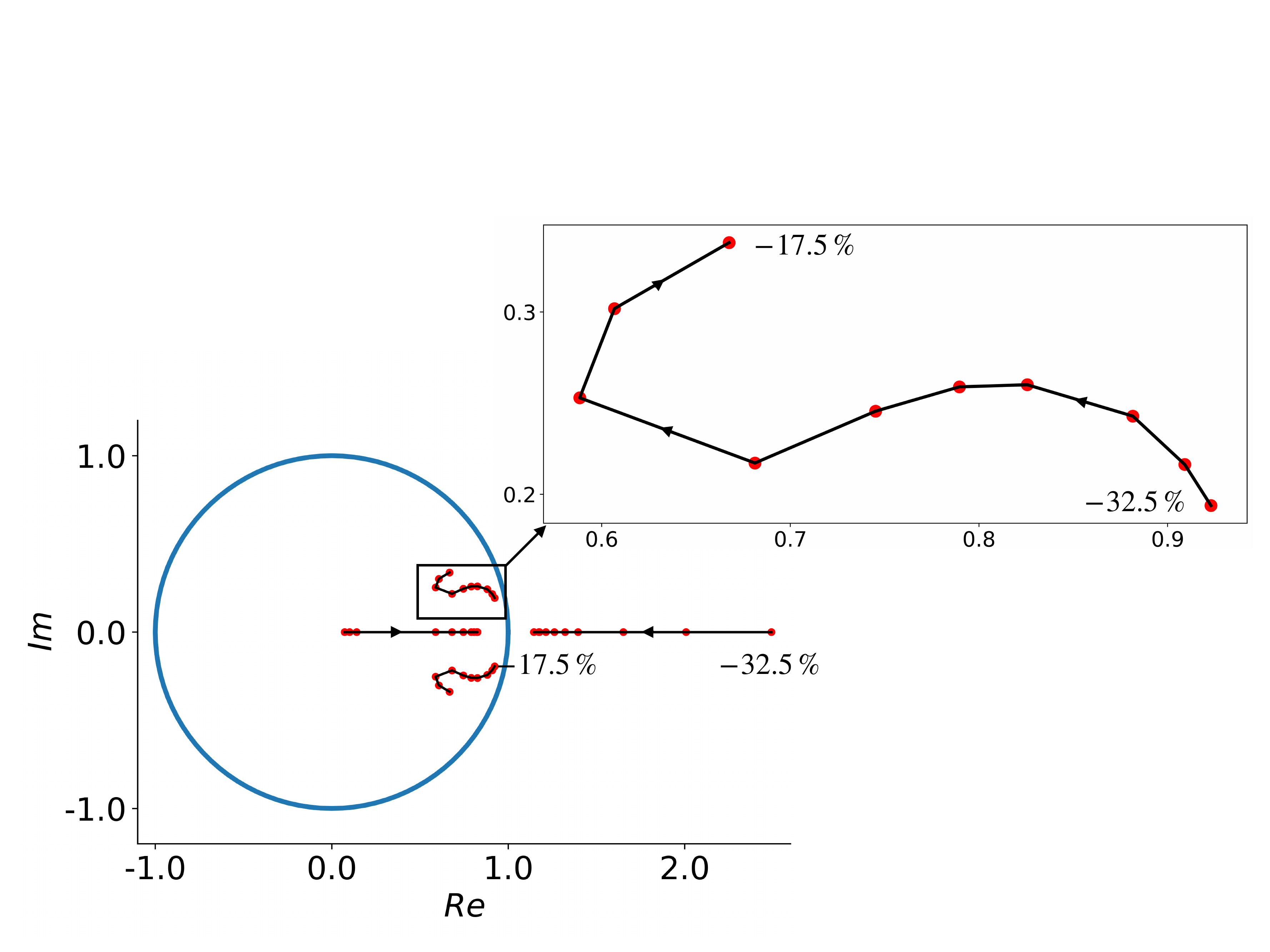}
\caption{Sensitivity of the eigenvalues with respect to the reciprocal position between wing-bone frame and the body frame.}
\label{fig:eigen_locus}
\end{figure}

Results are illustrated in Figure~\ref{fig:eigen_locus} where the locus of the eigenvalues is reported for different stability margins.
It shows that the eigenvalues locus (and therefore stability) is governed by the wing position, relative to the center of mass, and thus from the generation of an adequate pitching moment. Looking at Figure~\ref{fig:eigen_locus}, the expanding eigenvalue is smaller as the wing tends to get closer from the center of mass. This means that the capability of the wing to generate negative (nose-down) moment, is beneficial for the global stability behavior. In contrast, if the position of the wing is too much ahead from the center of mass, and consequently the capability of generating nose-down moment is reduced, the absolute value of the expanding eigenvalue, drastically increases, leading the system to be more and more unstable.\\
Thus the characteristic doubling time is putatively modulated by the wing kinematics and position, and is not unique for a given specie as suggested by~\cite{dietl2008}.\\
Importantly, none of the tested configuration, corresponds to a stable solution. The distance between the wing insertion point and the center of mass could be made smaller to continue bringing the largest eigenvalue close to the unit circle, but there is a threshold above which the wing is not capable of generating enough pitching up moment, for guaranteeing the existence of a limit cycle solution. Indeed, zero average pitching moment over one period is a necessary condition of existence of a limit cycle. In sum, we did not manage to find such limit cycles for $SM > -17.5\%$, and thus no stable limit cycle could be found.\\
%

\section{Conclusions}\label{sec:conclusions}
Flapping flight stability is a central concept for understanding how complex a control scheme is or needs to be in animal fliers, such as birds, or ornithopters. Experimental, theoretical, and numerical studies on such "flapping systems" have provided valuable insights on their dynamic and stability~(\cite{taylor2002}\cite{taylor2003}\cite{xiong2008}). With this contribution we have made a step forward by using a new model which is more accurate than existing ones in a couple of ways: (1) our wing is morphing during the cycle and we have enriched the bird motion by introducing critical degrees of freedom of the wing, especially the shoulder sweep angle; and (2) we have considered the wake effect on aerodynamic force production via a quasi-steady lifting line approach. 

Several trimmed trajectories have been retrieved by varying the gait parameters. In particular, the wingbeat amplitude relates quasi-linearly with the climbing angle and linear velocity in the explored range, thus determining whether the bird climbs, descends, or stays at constant altitude. It suggests that this degree of freedom could be a central control parameter for achieving level flight.

Furthermore, Floquet theory combined with multiple-shooting algorithms is confirmed as an elegant and powerful framework for analyzing the solutions of such flapping gaits. It turns out that the relative position between the wing and bird center of mass clearly affects pitching moments and global stability. Since our wings are massless, this effect has been investigated by moving the relative position between the center of mass and shoulder joint, whereas in reality this can be obtained by sweeping movements. Although the expanding eigenvalue gets closer to the unit circle when the center of mass approaches the wing root, stable configurations have not been found. In sum, the wing cannot generate a fully stabilizing effect in pitch. If so, birds would need to continuously rely on sensory feedback to adapt their gait via active control. Nevertheless, at least two important complementary aspects have not been investigated here. Stabilizing benefits could indeed arise either from wing compliance or from the aerodynamic contribution of the tail. These will be topic of our future investigations. 

Other questions regarding inertial contributions of bird wings and head to dynamics, also remain open. Comprehensive answers within this framework would require substantial modifications of the equations of motion adopted here.  

\newpage
\appendix
\section{Comparison between two methods for computing the Jacobian matrix}\label{appendix:jacobian}
\begin{figure}[!htbp]
\begin{center}
\begin{minipage}[c]{.45\textwidth}
\centering
\includegraphics[width=1.\textwidth]{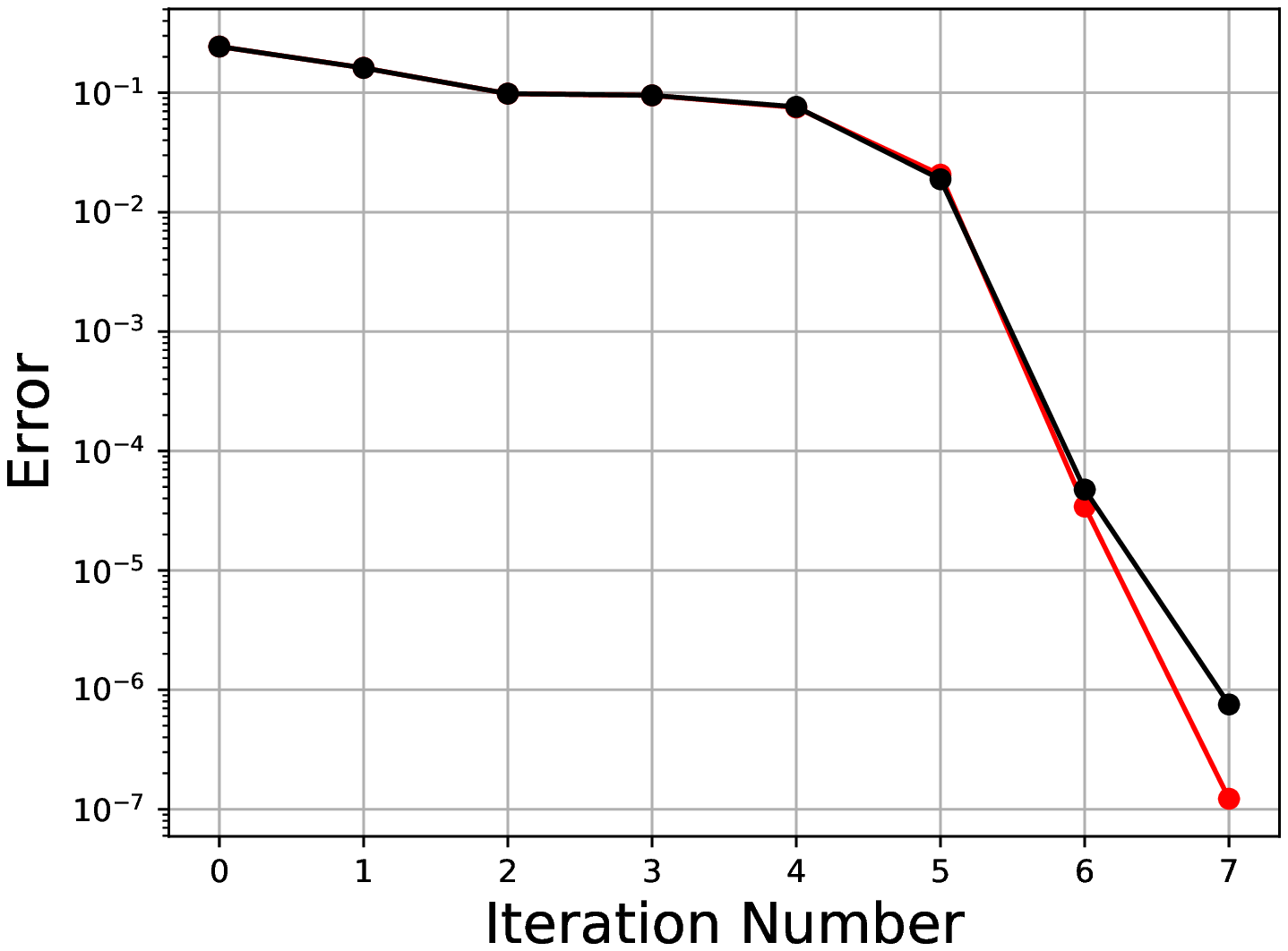}
\end{minipage}%
\hspace{10mm}%
\begin{minipage}[c]{.45\textwidth}
\centering
\includegraphics[width=1.\textwidth]{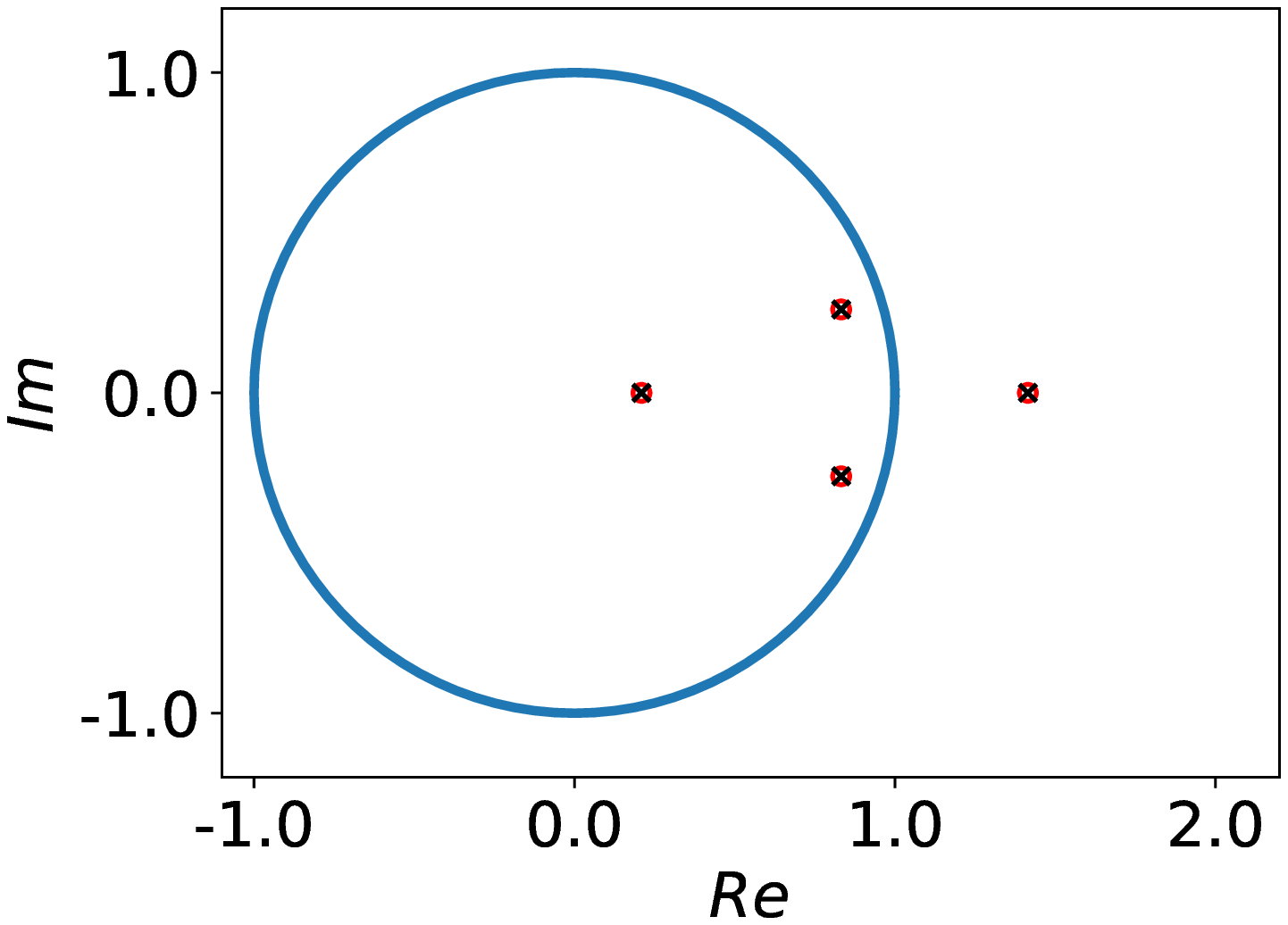}
\end{minipage}
\end{center}	
\caption{Left: Error evolution of the reference limit cycle solution, computed with a semi-analytical approach (black) and a numerical approach (red). Right: Floquet multipliers of the reference limit cycle computed with a semi-analytical approach (black) and a numerical approach (red).}
\label{fig:comp_jacobians}
\end{figure}

Section~\ref{subsec:jacobiancomputation} reported a so-called semi-analytical approach for computing the Jacobian matrix associated with the limit cycle. The method was called semi-analytical because, while it relied on a analytical derivation of the state-space matrix $A$, it still required a numerical estimation of the first partial derivatives of the aerodynamic forces. In this Appendix, we report a concurrent approach for computing the Jacobian matrix, that directly relies on a numerical estimation of its elements. More precisely, the numerical estimation of the Jacobian matrix relies on a finite difference approach, to numerically estimate the derivatives of the state-space flow. The main drawbacks of this numerical approach are its computational cost, and an accurate choice of the integration step size if the system (\ref{eqn:jacobianreshaped_formula}) is too stiff.\\ 
In this method, the $i,j$ component of the Jacobian matrix is computed by a finite differentiation of the perturbed trajectory along each state variable, i.e.
\begin{equation}
	\mathbb{J}_{i,j} (\mathbf{x}_0) \Big \rvert_{t}^{t+T} = \frac{f_{i}(\mathbf{x_0} + \varepsilon\mathbf{\hat{e}}_{j}) \Big \rvert_{t}^{t+T} -f_{i}(\mathbf{x_0}) \Big \rvert_{t}^{t+T} }{\varepsilon} 
\label{eqn:jac_numerical}
\end{equation}
The same reference case of Table~\ref{tab:parameters} has been used for comparing both methods for computing the Jacobian matrix, i.e. solving Equation~\ref{eqn:jacobianreshaped_formula} and Equation~\ref{eqn:jac_numerical} respectively. The step size of the numerical integration was set constant and equal to $dt=T/100$.\\
The computational time per iteration with the semi-analytical approach was found to be about $40s$ while the numerical approach took more than than the doube, around $95s$ per iteration. A comparison of the resulting errors and Floquet multipliers are provided in Figure~\ref{fig:comp_jacobians}.
In conclusion, both approaches provided the same eigenvalues with a very similar convergence rate, although the semi-analytical approach took less than half of the computational time of the fully numerical one.

\section*{Acknowledgement}
This work was supported by Fédération Wallonie-Bruxelles (FWB) under the Action de recherche concertée (ARC) RevealFlight (grant number \textbf{17/22-080, REVEALFLIGHT} – The reverse- engineering of flight: a bottom-up reproduction of bird biomechanics and of self-organization into a flock).\\
We acknowledge Pierre-Antoine Absil, for the precious advices for the implementation of the multiple-shooting algorithm and Tommaso Sartor for helping in stucturing the code.

\bibliography{References}

\begin{thebibliography}{10}
\newcommand{\enquote}[1]{``#1''}

\bibitem{shyy1999}
Shyy, W., Berg, M., and Ljungqvist, D., \enquote{Flapping and flexible wings
  for biological and micro air vehicles,} {\em Progress in aerospace
  sciences\/}, Vol.~35, No.~5, 1999, pp.~455--505.

\bibitem{taylor2002}
Taylor, G. and Thomas, A., \enquote{Animal flight dynamics II. Longitudinal
  stability in flapping flight,} {\em Journal of theoretical biology\/},
  Vol.~214, No.~3, 2002, pp.~351--370.

\bibitem{taylor2003}
Taylor, G.~K. and Thomas, A.~L., \enquote{Dynamic flight stability in the
  desert locust Schistocerca gregaria,} {\em Journal of Experimental
  Biology\/}, Vol.~206, No.~16, 2003, pp.~2803--2829.

\bibitem{xiong2008}
Xiong, Y. and Sun, M., \enquote{Dynamic flight stability of a bumblebee in
  forward flight,} {\em Acta Mechanica Sinica\/}, Vol.~24, No.~1, 2008,
  pp.~25--36.

\bibitem{iosilevskii2014b}
Iosilevskii, G., \enquote{Forward flight of birds revisited. Part 2: short-term
  dynamic stability and trim,} {\em Royal Society open science\/}, Vol.~1,
  No.~2, 2014, pp.~140249.

\bibitem{taylor2005}
Taylor, G.~K. and {\.Z}bikowski, R., \enquote{Nonlinear time-periodic models of
  the longitudinal flight dynamics of desert locusts Schistocerca gregaria,}
  {\em Journal of the Royal Society Interface\/}, Vol.~2, No.~3, 2005,
  pp.~197--221.

\bibitem{dietl2008}
Dietl, J.~M. and Garcia, E., \enquote{Stability in ornithopter longitudinal
  flight dynamics,} {\em Journal of Guidance, Control, and Dynamics\/},
  Vol.~31, No.~4, 2008, pp.~1157--1163.

\bibitem{vanderberg1995}
Berg, C. and Rayner, J., \enquote{The moment of inertia of bird wings and the
  inertial power requirement for flapping flight,} {\em Journal of experimental
  biology\/}, Vol.~198, No.~8, 1995, pp.~1655--1664.

\bibitem{dietl2011}
Dietl, J., Herrmann, T., Reich, G., and Garcia, E., \enquote{Dynamic modeling,
  testing, and stability analysis of an ornithoptic blimp,} {\em Journal of
  Bionic Engineering\/}, Vol.~8, No.~4, 2011, pp.~375--386.

\bibitem{taha2014}
Taha, H.~E., Hajj, M.~R., and Nayfeh, A.~H., \enquote{Longitudinal flight
  dynamics of hovering MAVs/insects,} {\em Journal of Guidance, Control, and
  Dynamics\/}, Vol.~37, No.~3, 2014, pp.~970--979.

\bibitem{etkin1959}
Etkin, B. and Reid, L.~D., {\em Dynamics of flight\/}, Vol.~2, Wiley New York,
  1959.

\bibitem{casarosa2013}
Casarosa, C., {\em Meccanica del volo\/}, Didattica e Ricerca. Manuali, Pisa
  University Press, 2013.

\bibitem{taha2012}
Taha, H.~E., Hajj, M.~R., and Nayfeh, A.~H., \enquote{Flight dynamics and
  control of flapping-wing MAVs: a review,} {\em Nonlinear Dynamics\/},
  Vol.~70, No.~2, 2012, pp.~907--939.

\bibitem{colognesi2020}
Colognesi, V., Ronsse, R., and Chatelain, P., \enquote{A model coupling
  biomechanics and aerodynamics for high-fidelity simulations of controlled
  flapping flight,} {\em Bioinspiration \& Biomimetics\/}, 2020.

\bibitem{Wu:2003}
Wu, J.-c. and Popovi{\'c}, Z., \enquote{Realistic Modeling of Bird Flight
  Animations,} {\em ACM Transactions on Graphics (TOG)\/}, Vol.~22, No.~3, July
  2003, pp.~888--895.

\bibitem{buresti2012}
Buresti, G., {\em Elements of fluid dynamics\/}, Vol.~3, World Scientific
  Publishing Company, 2012.

\bibitem{floquet1883}
Floquet, G., \enquote{Sur les {\'e}quations diff{\'e}rentielles lin{\'e}aires
  {\`a} coefficients p{\'e}riodiques,} {\em Annales scientifiques de
  l'{\'E}cole normale sup{\'e}rieure\/}, Vol.~12, 1883, pp. 47--88.

\bibitem{ChaosBook}
Cvitanovi\'c, P., Artuso, R., Mainieri, R., Tanner, G., and Vattay, G., {\em
  \em Chaos: Classical and Quantum\/}, Niels Bohr Institute, Copenhagen 2016.

\bibitem{strogatz2018}
Strogatz, S.~H., {\em Nonlinear Dynamics and Chaos with Student Solutions
  Manual: With Applications to Physics, Biology, Chemistry, and Engineering\/},
  CRC Press, 2018.

\bibitem{seydel2009}
Seydel, R., {\em Practical bifurcation and stability analysis\/}, Vol.~5,
  Springer Science \& Business Media, 2009.

\bibitem{lust2001}
Lust, K., \enquote{Improved numerical Floquet multipliers,} {\em International
  Journal of Bifurcation and Chaos\/}, Vol.~11, No.~09, 2001, pp.~2389--2410.

\bibitem{keller1968}
Keller, H., {\em Numerical Methods for Two-point Boundary-value Problems\/}, A
  Blaisdell Book in Numerical Analysis and Computer Science, Blaisdel, 1968.

\bibitem{quarteroni2010}
Quarteroni, A., Sacco, R., and Saleri, F., {\em Numerical mathematics\/},
  Vol.~37, Springer Science \& Business Media, 2010.

\bibitem{marquardt1963}
Marquardt, D.~W., \enquote{An algorithm for least-squares estimation of
  nonlinear parameters,} {\em Journal of the society for Industrial and Applied
  Mathematics\/}, Vol.~11, No.~2, 1963, pp.~431--441.

\bibitem{dednam2015}
Dednam, W. and Botha, A.~E., \enquote{Optimized shooting method for finding
  periodic orbits of nonlinear dynamical systems,} {\em Engineering with
  Computers\/}, Vol.~31, No.~4, 2015, pp.~749--762.

\bibitem{tobalske1996}
Tobalske, B. and Dial, K., \enquote{Flight kinematics of black-billed magpies
  and pigeons over a wide range of speeds,} {\em Journal of Experimental
  Biology\/}, Vol.~199, No.~2, 1996, pp.~263--280.

\bibitem{Taylor2006}
Taylor, G., Bomphrey, R., and 't~Hoen, J., \enquote{Insect flight dynamics and
  control,} {\em 44th AIAA Aerospace Sciences Meeting and Exhibit\/}, 2006,
  p.~32.

\end{thebibliography}
\bibliographystyle{aiaa}
\end{document}